\documentclass[preprint,12pt,authoryear]{elsarticle}

\usepackage{amssymb}
\usepackage{amsthm}
\usepackage{amsmath}
\usepackage{amsfonts}
\usepackage{bm}
\usepackage[flushleft]{threeparttable}
\usepackage{multirow}
\usepackage{booktabs}
\usepackage{gensymb}
\usepackage{makecell}
\newtheorem{myDef}{Definition}
\newtheorem{proposition}{Proposition}

\usepackage{lineno}
\journal{arXiv}

\begin{document}

\begin{frontmatter}



\title{Baumol's climate disease}


\author[label1,label2]{Fangzhi Wang}
\author[label1,label2]{Hua Liao}
\author[label3,label4,label5,label6,label7,label8,label9]{Richard S.J. Tol}

\affiliation[label1]{organization={School of Management and Economics, Beijing Institute of Technology},%Department and Organization
            city={Beijing},
            postcode={100081}, 
            country={China}}

\affiliation[label2]{organization={Center for Energy and Environmental Policy Research, Beijing Institute of Technology},%Department and Organization
            city={Beijing},
            postcode={100081}, 
            country={China}}

\affiliation[label3]{organization={Department of Economics, University of Sussex}, 
            city={Falmer},
            postcode={BN1 9RH}, 
            country={UK}}

\affiliation[label4]{organization={Institute for Environmental Studies, Vrije Universiteit},
            city={Amsterdam},
            country={the Netherlands}}

\affiliation[label5]{organization={Department of Spatial Economics, Vrije Universiteit},
            city={Amsterdam},
            country={the Netherlands}}

\affiliation[label6]{organization={Tinbergen Institute}, 
            city={Amsterdam},
            country={the Netherlands}}

\affiliation[label7]{organization={CESifo},
            city={Munich},
            country={Germany}}

\affiliation[label8]{organization={Payne Institute for Public Policy, Colorado School of Mines},
            city={Golden},
            state={CO},
            country={USA}}

\affiliation[label9]{organization={College of Business, Abu Dhabi University},
            city={Abu Dhabi},
            country={UAE}}

\begin{abstract}
We investigate optimal carbon abatement in a dynamic general equilibrium climate-economy model with endogenous structural change.  By differentiating the production of investment from consumption, we show that social cost of carbon can be conceived as a reduction in physical capital. In addition, we distinguish two final sectors in terms of productivity growth and climate vulnerability. We theoretically show that heterogeneous climate vulnerability results in a climate-induced version of Baumol's cost disease. Further, if climate-vulnerable sectors have high (low) productivity growth, climate impact can either ameliorate (aggravate) the Baumol's cost disease, call for less (more) stringent climate policy. We conclude that carbon abatement should not only factor in unpriced climate capital, but also be tailored to Baumol's cost and climate diseases.
\end{abstract}


%\item A dynamic general equilibrium climate-economy model with endogenous structural change
%\item Denominate the social cost of carbon in terms of investment, attributing climate change to disinvestment rather than malconsumption
%\item Heterogeneous climate vulnerability between sectors leads to Baumol's cost disease
%\item Aggregate climate impact is exacerbated if the more climate-vulnerable sectors have slow technological progress


\end{frontmatter}

\newpage
\section{Introduction}
\label{sec:Introduction}

Economic growth may be hampered by a stagnant services sector \citep{baumol1967macroeconomics} and by climate change \citep{nordhaus1996regional}. This paper shows that climate change would have a similar effect as Baumol's cost disease. Optimal greenhouse gas emission reduction should take this into account. 

The paper can be summarized with three arguments. First, climate proxied by temperature rise is an \emph{essential capital 'bad'} used in economic production\footnote{There are few precedents for considering climate as capital \citep{arrow2004we,10.2307/26265518,barrage2020optimal}. See also \citet{NBERw22060} who viewed environmental quality as a stock of capital. Likewise, one may hence interpret climate capital as the desirability of the climate conditions, which can be represented by a function of temperature. Then, climate capital is a capital good. However, in line with the current literature commonly using temperature in the production function \citep[e.g.][]{nordhaus2017revisiting,golosov2014optimal,barrage2020optimal}, we use temperature here to proxy climate capital directly. This manner may be counterintuitive at first sight, but can provide modelling convenience.}. Climate is an \emph{essential} input to economic production. Growing crops requires decent climate conditions including appropriate temperature, humidity, sunshine, etc. White-collar workers in high-tech companies demand air-conditioning. A higher temperature is typically found to exert an adverse impact on economic production \citep{Tol2018}. In this sense, climate is a \emph{'bad'} capital. Climate is a \emph{stock}. Climate change is driven by the historical accumulation of emissions. Thus, climate embodies similar inter-temporal properties as physical or human capital \citep{lucas1988mechanics}. 

Because climate is underpriced, carbon dioxide is overinvested. When climate is recognized as capital bad, emission of carbon dioxide is naturally a disinvestment. In addition, climate is a public good, as climate capital is both \emph{nonrival} and \emph{nonexcludable}. On one hand, like ideas, climate capital is \emph{nonrival}. Once the cost of attaining a lower level of climate capital, \textit{e.g.} temperature, has been incurred, the climate capital can be leveraged repeatedly at no additional costs. One agent's use of temperature does not affect another agent's use. Furthermore, climate capital is \emph{nonexcludable}, unlike (patented, trademarked or copyrighted) ideas. Nobody is capable of appropriating the property of climate, whether they are individuals, firms, or countries. No agent can stop another agent from using temperature. This implies that there are endless incentives to invest carbon dioxide emissions in the atmosphere. In plain language, because emitting carbon dioxide is free, profit-maximizing firms do not care whether carbon emissions are contributing to climate change.

The second argument is that heterogeneous climate vulnerability between sectors reflects the potential to produce with climate capital, causing the climate version of Baumol's cost disease\textemdash \emph{Baumol's climate disease} in short. Although climate capital is uniformly acquired, the ability to produce varies between sectors at any level of climate capital. For some sectors, climate capital is by its very nature a core requisite, implying a higher climate vulnerability. Other economic activities are not fastidious about climate conditions. As climate change progresses due to overinvested carbon, the climate vulnerability gap can be expected to widen between these sectors. This division echoes  \citet{baumol1967macroeconomics} that conceptually differentiated progressive with non-progressive sectors due to their technological structure. If products in sectors are not substitutes, more and more productive factors would flow from progressive to stagnant sectors. Eventually, the overall productivity growth declines as the non-progressive sectors are expanding, notoriously known as Baumol's cost disease\footnote{\citet{nordhaus2008baumol} provides empirical evidence for Baumol's cost disease. Some recent studies explore Baumol's cost disease in dynamic growth models, and confirm that the rising price of services relative to goods will slow aggregate productivity growth \citep{ngai2007structural,herrendorf2021structural}.}. Likewise, if a sector is progressively damaged by climate change, productive factors are increasingly absorbed into it, rendering the overall economy more vulnerable to climate change. In effect, higher climate vulnerability is analogous to slower productivity growth, giving rise to Baumol's climate disease.

The third argument is that evaluating climate impact should factor in technology-driven structural change. Baumol's climate disease recognizes the role of heterogeneous climate vulnerability, but its net effect depends on both technological structure and climate vulnerability. More specifically, climate vulnerability in each sector can be either aggravated or compensated by technological change. If a sector with high climate vulnerability is blessed with high technological growth, climate impact can be less worrisome because the technological structure makes this sector more resilient. On the contrary, if a sector with high climate vulnerability is further depressed by gloomy technological prospects, the relative price of production in this sector will grow even higher. Consequently, Baumol's cost disease bites harder. In this case, reducing carbon emissions can generate dual benefits including both avoided climate damages and moderated cost disease.

Although the economics literature on climate impact is burgeoning \citep[e.g.][etc.]{NBERw24893, carleton2022valuing, waldinger2022economic, 10.1093/restud/rdad042, acemoglu2023mirage, Tol2023}, the conception of climate as capital, to our knowledge, has not been formalized in climate-economy models. The paper makes some progress in this regard. We start with standard properties that naturally come along with identifying climate as a capital. For example, climate capital delivers negative returns, and the marginal return to climate capital is non-decreasing at a higher temperature. Besides, nonrival climate capital features decreasing returns to scale. Because climate capital is also nonexcludable, the nonrivalry contributes to the negative climate externality, in contrast to the positive externality in the knowledge capital. \citet{Nobel2018} points out that externalities bridge the contributions of Romer and Nordhaus. Along its route, we emphasize that nonrivalry and nonexcluability of climate capital are the fundamental cause of climate change, and that integrated assessment models should assume non-constant-return-to-scale technology for climate capital in production functions. Moreover, the paper interprets climate damage functions as the ability to produce with climate capital, which can differ in sectors.

To shed light on optimal carbon abatement under structural change, this paper then establishes a dynamic general equilibrium climate-economy model with endogenous structural change. The climate-economy linkage stands on the shoulders of \citet{golosov2014optimal}, \citet{nordhaus2017revisiting} and \citet{barrage2020optimal}. Economic production requires energy as essential input, which will generate carbon emissions. Unabated carbon emissions will enter the atmosphere and affect climate capital that is indispensable for economic activities. On the one hand, our model differentiates \emph{two} final production sectors, both of which use physical capital, labor, energy, knowledge and climate as input. In addition, both sectors adopt non-constant-return-to-scale technology when using knowledge and climate for production. Thus, technological growth and climate vulnerability combine to shape the relative price between two sectors. On the other hand, production in two sectors are required for both consumption and investment (\emph{two} final expenditures). Therefore, the model is characterized by a \emph{two-by-two} structure.  Following \citet{herrendorf2021structural}, \citet{garcia2021investment} and \citet{NBERw27731}, we induce structural transformation in both consumption and investment via the price effect\footnote{The income effect is the other important force in incurring structural change \citep{buera2009can,boppart2014structural,comin2021structural,alder2022theory}. }. For both expenditures, the substitution elasticity is less than unity between two final products \citep{herrendorf2013two, herrendorf2021structural, garcia2021investment}. Hence, the comparative scarcity between two products determines their relative price, leading up to structural change. The results of the model are generalized as follows.

First, we provide a novel representation of the social cost of carbon based on investment. By differentiating the production of investment and consumption, we find that the trade-off between abatement costs and avoided climate damage is based on investment rather than consumption. This result theoretically supports that climate is a capital bad. Thus, the real cost of carbon can be conceived as a drag on physical capital. Existing studies usually define the social cost of carbon as the consumption loss due to an additional tonne of CO\textsubscript{2} \citep{nordhaus2014estimates,golosov2014optimal,barrage2020optimal}. In one-sector growth model, economic production is utilized for either consumption or investment without differences. Thus, investment-equivalent social cost of carbon is identical to consumption-equivalent social cost of carbon therein. Social cost of carbon denominated in terms of consumption has straightforward welfare implications because consumption is a direct measure of welfare, unlike investment. By comparison, \citet{10.2307/26265518} adopts social cost of carbon to approximate the change in environmental capital, implying a cost in investment. In their language, the social cost of carbon denominated by investment reflects the cost in wealth instead of directly in income, and hence also has welfare implications owing to its close relevance to sustainability. It should be noted that the choice of numeraire only influences the numerical value of social cost of carbon, whereas optimal allocation is immune to either choice. Investment-equivalent social cost of carbon is associated with consumption-equivalent social cost of carbon by the relative price of investment to consumption.

Second, we theoretically demonstrate how climate change can influence Baumol's cost disease. For one thing, we assume the Cobb-Douglas functions in two sectors with the same factor intensity. For another, we allow for different technological growth and climate vulnerability in both sectors. Absent different technological growths, the relative price in the more climate-vulnerable sector will increase as climate change proceeds, because production in this sector is comparatively scarcer than that of less climate-vulnerable sector, exacerbating the climate impact via Baumol's climate disease. Accounting for heterogeneous technological growth, the relative price between two sectors is pinned down by the relative technological growth combined with relative climate vulnerability. Suppose that the impact of technological growth dominates that of climate vulnerability. The relative price between two sectors will increase further when the more susceptible sector is also experiencing slower technological growth. That is, climate change exacerbates the Baumol's cost disease. By comparison, when the more susceptible sector has high productivity growth, the relative price will still go up, but to an extent less than absent climate impact. In other words, although climate change takes a toll in both sectors, it ameliorates the Baumol's cost disease.

Third, we quantify the model. For expositional convenience, we devide the economy into goods and services. Dividing the economy into goods and services is not uncommon in the structural change literature \citep[e.g.][]{moro2015structural,leon2020rise,herrendorf2021structural}, motivated by the observed slower technological growth in services. In addition, we find some preliminary empirical evidence that services are less susceptible to climate change than goods.  In line with the literature, we assume that the productivity growth in the goods sector is three times that in the services sector. Absent climate damage, the capital stock in 2100 is set identical to that in the DICE model for comparability. We adopt the damage function from DICE, assuming that the relative damage level in the goods sector is two times higher than in services. Using the data from World Bank, we pin down the factor shares used in each sector in the initial period.

Fourth, our numerical results validate that Baumol's cost disease is an important consideration for carbon abatement policy. When two sectors are only different in climate vulnerability, capital stock is reduced by 12.34\% in 2100 and 19.37\% in 2150, compared to a decrease of 11.25\% and 17.47\% under homogeneous climate vulnerability. Thus, Baumol's climate disease aggravates  aggregate climate impact. As a consequence, a more stringent climate policy is required to achieve the optimal allocation, which achieves the net-zero carbon emission in 2095, earlier than in 2100 for homogeneous climate vulnerability. When also accounting for differentiated productivity growth in the baseline model, there is little difference for capital stock between heterogeneous or homogeneous climate vulnerability. Moreover, as consumption responds to climate with lags, consumption is even improved under heterogeneous climate vulnerability in the considered periods. This is because the expanding services sector is less vulnerable to climate change, or alternatively, the goods sector that is more vulnerable enjoys a higher productivity growth, showing higher resilience to climate change. A less strict abatement policy is required for optimal allocation, putting off 10 years for achieving net-zero emissions. Furthermore, we consider a counterfactual scenario where the services sector is assumed to be more climate-vulnerable. Quantitative results come to that the climate damage on capital stock will be aggravated by a further loss of 9.20 percentage points if climate change increases the relative price of services, \emph{i.e.} the more severe Baumol's cost disease. Accordingly, net-zero emissions are required to be achieved twenty years earlier.

In addition, we quantify two definitions of the social cost of carbon. In the presence of optimal abatement, the social cost of carbon at 2100 stands for an investment loss of \$254 per tone of CO\textsubscript{2}, and a consumption loss of \$182 per tone of CO\textsubscript{2}. Compared to investment, consumption is composed of a higher ratio of services. As the relative price of services is climbing over time, so will the price of consumption relative to investment. Given a numeraire with a higher price, consumption-equivalent social cost of carbon is lower than investment-equivalent one. Moreover, we find that a lower social time preference rate will generate a larger gap between the values of two definitions.

Although we only simulate goods and services in the model, our results imply that climate policy should be tailored to Baumol's joint cost-and-climate disease. The more sectors are there in the economy with low productivity growth and high climate vulnerability, \emph{i.e.} the more acute the Baumol's joint cost-and-climate disease, the stronger incentives to reduce the carbon emissions. Otherwise, the long-run economy growth would be plagued by both climate change and exacerbated Baumol's cost disease. In addition, the social planner need increase adaptation investment into any sector with both high climate vulnerability and low productivity growth. Thus, our paper has policy implications for both abatement and adaptation under structural change.

This paper relates to the current literature in the following ways. First, existing literature typically denominates the social cost of carbon in terms of consumption \citep{nordhaus2014estimates, golosov2014optimal}, we propose an alternative definition in term of investment. In so doing, we show that social cost of carbon is a drag on productive capital rather than directly on consumption. We also demonstrate two definitions of social cost of carbon can be bridged by the relative price of investment to consumption. Second, the paper is relevant to studies on climate damages. \citet{casey2021understanding} assume that the climate system evolves exogenously, and analyzes the climate damages on consumption and investment with the focus on heterogeneous damages between sectors. Our study is consistent with theirs in finding that heterogeneous climate vulnerability will exacerbate aggregate damage level, which could be explained by the less than unity substitution elasticity between each product in producing investment and consumption. This paper, however, differs from theirs in that we focus on the background of structural change. Baumol's cost and climate diseases are critical factors in determining the realized impact of climate change. Third, the paper falls within the broad category of structural change economics \citep{herrendorf2014growth}, and we add that climate vulnerability is also complementary to incurring the price effect.

The remainder of the paper is organized as follows. Section \ref{sec:Climate capital} formalize climate as a capital. Section \ref{sec:Model and theory} establishes a climate-economy model and theoretically analyzes the incentives to disinvest into climate capital. Section \ref{sec:Calibration} introduces the calibration process. Section \ref{sec:Quantitative results} discusses the quantitative results. Some sensitivity analyses and extensions are included in Section \ref{sec:Sensitivity and extensions}, and Section \ref{sec:Conclusion} concludes the paper.

\section{Climate capital in production function}
\label{sec:Climate capital}
This section discusses several fundamental properties associated with climate capital, which have not been formalized previously. Where necessary, we compare climate capital with other common factors in production. In so doing, we are able to interpret the cause of climate change from the perspective of investment. In addition, we note the link between climate capital and climate damage function that is commonly used in climate economics literature.

Temperature change $T_t$ is considered as an appropriate proxy for climate capital, consistent with extant studies \citep{dice2023,golosov2014optimal,10.1093/restud/rdad042}. In addition to climate capital, economic production at period $t$ also requires physical capital $K_t$, labor $L_t$, ideas $A_t$, and energy $E_t$.
\begin{equation}
\label{eq:prod}
    Y_t=F(A_t,T_t,K_t,L_t,E_t)
\end{equation}
where $F$ represents some technology to utilize these factors in production. The last argument energy $E_t$ can be either fossil fuels that will emit carbon dioxide or renewable energy with no carbon emissions.

\subsection{Negative and increasing returns}

Unlike other factors, climate represented by global mean temperature change is a bad capital. More input of other factors (\emph{i.e.} physical capital, human capital, energy and ideas) generate more revenues, whereas an increase in temperature causes economic losses:
\begin{equation}
    \frac{\partial F}{\partial T_t} < 0
\end{equation}
Moreover, as the temperature increases further, the marginal returns to climate capital is increasing:
\begin{equation}
    \frac{\partial^2 F}{\partial^2 T_t} > 0
\end{equation}
Increasing returns to climate capital captures both the beliefs \citep{dice2023,weitzman2010damages,pindyck2021we} and some empirical evidence \citep{burke2015global,newell2021gdp} that climate impact is exacerbated at a higher temperature.

\subsection{Decreasing returns to scale}
In nature, climate capital is a \emph{public good}, because it is both nonrival and nonexcludable. Consequently, climate capital is overinvested, giving rise to global warming.

Climate capital is nonrival. An individual's use of climate capital does not preclude others from using climate capital. Once climate capital is produced (represented by global temperature change), all firms use it for economic production without paying for additional costs. When global mean temperature rises, an enduring heat wave may ensue in an African village, devastating the harvests of all peasants. One peasant's adversity cannot lower the possibility or extent of another peasant in the same village. In Singapore, a coastal high-tech company is simultaneously plagued by the sea-level rise that floods the working office due to the same climate capital, \emph{i.e.} global mean temperature change. While the sea level is rising in Singapore, miserable peasants find no reasons to believe that flooding makes global temperature increase or decrease, nor will the company think the heat wave in Africa will cool down the global suddenly. Climate capital, once produced, can be enjoyed by everyone and hence is not scare. However, good climate capital is scare in that, for example, there is some unknown level of global temperature change leading up to maximum global economic production, \emph{ceteris paribus}. 

Nonrivalry of climate capital implies that production is characterized by decreasing returns to scale. The standard replication argument is valid for physical capital, human capital and energy, but not for ideas and climate capital. \citet{romer1986increasing,romer1990endogenous} has illuminated that ideas are nonrival and consequently that economic production features increasing returns to scale. Because climate is a bad capital, climate capital is characterized by decreasing returns to scale. In other words, for any $\lambda > 1$, we have:
\begin{align}
    F(A_t,T_t,\lambda K_t,\lambda L_t,\lambda E_t) &= \lambda F(A_t,T_t,K_t,L_t,E_t)\notag\\
    F(\lambda A_t,T_t,\lambda K_t,\lambda L_t,\lambda E_t) &> \lambda F(A_t,T_t,K_t,L_t,E_t)\notag\\
    F(A_t,\lambda  T_t,\lambda K_t,\lambda L_t,\lambda E_t) &< \lambda F(A_t,T_t,K_t,L_t,E_t)
\end{align}

Climate capital is also nonexcludable, leading up to overinvested carbon. Compared to knowledge capital that can be partially excludable in the presence of patents, the property of climate capital cannot be appropriated in reality. Suppose that a firm reduces one unit of carbon investment into climate capital and shoulders the associated abatement costs. Because climate capital is not excludable, all firms in the economy can benefit from this one unit of reduced carbon that lowers the level of climate capital. Thus, the abatement costs for reduced carbon cannot be compensated for by private revenues. In the market, when the social price of carbon is not defined, no firm is motivated to disinvest carbon into climate capital.

\subsection{Climate capital and damage function}
Although climate capital proxied by global mean temperature is uniform to everybody, the ability to produce with climate capital can be quite different. In other words, climate impact is heterogeneous. This can be explained by geographic endowments, adaptation technology, industry structure, etc. In existing literature \citep[e.g.:][]{dice2023,golosov2014optimal,barrage2020optimal,10.1093/restud/rdad042}, the damage function pioneered by \citep{nordhaus1992optimal} is commonly leveraged to reflect the productivity of climate capital. Admittedly, the damage function is among the most uncertain parts in climate-economy models in terms of both forms and parameters \citep{pindyck2021we}. The paper makes no efforts to determine any well-suited damage function. Instead, as we will see below, the analysis only requires that climate damage functions should be sector-specific so as to reflect the differentiated productivity of climate capital between sectors.

\section{Model and theory}
\label{sec:Model and theory}
This section establishes a dynamic climate-economy model with endogenous structural change. The climate-economy structure is borrowed from \citet{nordhaus2017revisiting}, \citet{golosov2014optimal} and \citet{barrage2020optimal}. The economic module is augmented with a two-by-two structure. On the one hand, we allow for two final production sectors with heterogeneous productivity growth rate and climate vulnerability, both of which can affect the relative price of products in two sectors\footnote{Our model can be easily extended to a multi-sector case.}. On the other hand, following \citet{greenwood1997long}, \citet{herrendorf2014growth} and \citet{foerster2022aggregate}, we differentiate two final expenditures of economic production--consumption and investment in terms of their compositions of final products. Thus, structural change can take place within both consumption and investment when the price in one sector changes relative to the other. We start from the competitive market and show its equivalence to social planner problem by introducing a carbon tax. 

Given the established model, we first theoretically show how to achieve the optimal allocation in the presence of climate externality. This result constitutes the fundamental incentive for addressing climate change--unpriced climate capital and overinvested carbon. Then, we show the relative price between two products, which can be perceived as the acuteness of the Baumol's cost disease. Reducing carbon changes the relative price, and hence becomes another important consideration in optimal carbon abatement decision.

\subsection{Households}

The economy accommodates an infinitely-lived, representative household whose lifetime utility is determined by:
\begin{align}
\label{eq:utility}
U_{0} \equiv \sum_{t=0}^{\infty} \beta^t U(C_{t})
\end{align}
where $\beta$ denotes the social time of preference and $C_t$ market consumption at period $t$. In fact, the household's utility can also be affected directly by climate change. For example, it can influence amenities, biodiversity or health conditions that households value\footnote{This consideration can be represented by either introducing a climate variable such as temperature rise into the utility function \citep{barrage2020optimal} or considering the climate impact on non-market goods \citep{Tol1994enpol, doi:10.1093/reep/rem024,drupp2021relative}.}. Since the paper focuses on economic dynamics, we abstract from these possibilities.

The consumption bundle is composed of two different products: 
\begin{align}
\label{eq:consumption production}
C_t = F_{C}(C_{1t},C_{2t})
\end{align}
where $C_{it}$ stands for consumption of product $i\in\{1,2\}$ respectively. $F_C$ captures the household's preference for each product, and can be regarded as a costless technology aggregating both consumption. 

The representative household should satisfy its budget constraint in each period:
\begin{align}
\label{eq:budget constraint}
p_{1t}C_{1t} + p_{2t}C_{2t} + K_{t+1} \leq w_t L_t + (1+r_t-\delta)K_t + \Pi_t + Tran_t
\end{align}
where $K_{t+1}$ is the capital holdings at time $t+1$, $w_t$ the wage rate, $L_t$ the labor supply, $r_t$ the rental rate of capital, $\delta$ the capital depreciation rate, $\Pi_t$ the dividends from the energy sector, $Tran_t$ the lump-sum transfer of carbon tax levied by the government. We set investment as the numeraire, normalizing its price to one in each period, and define $p_{it}$ as the price of product $i$.

Thus, the household’s first-order condition requires that saving decisions and allocations of goods and services in consumption must satisfy:
\begin{align}
\frac{U_{C1t}/p_{1t}}{U_{C1,t+1}/p_{1,t+1}}
=\frac{U_{C2t}/p_{2t}}{U_{C2,t+1}/p_{2,t+1}}=\beta(1+r_t-\delta)
\end{align}
where $U_{cit}$ represents the partial derivative of instantaneous utility function with respect to consumption $i$ at time $t$. This equation demonstrates that, between two subsequent periods, the representative household will equate the price-adjusted marginal rates of substitution in each consumption to the return rate on saving.

\subsection{Two final production sectors}
There are two final sectors whose production functions follow from Eq.(\ref{eq:prod}). Further, each sector $i\in\{1,2\}$ adopts a constant-returns-to-scale technology $\tilde{F}$ to combine physical capital, labor and energy, and satisfies the Inada conditions. By comparison, both climate capital and knowledge capital feature non-constant returns to scale, represented by $\hat{F}$. Thus, we have:
\begin{align}
         \label{eq:goods production}
    Y_{1t} &= F_1(A_{1t},T_t,K_{1t},L_{1t},E_{1t})\notag\\
           &=\hat{F_1}(A_{1t},T_t) \tilde{F_1}(K_{1t},L_{1t}, E_{1t})\\
         \label{eq:services production}
    Y_{2t} &= F_2(A_{2t},T_t,K_{2t},L_{2t},E_{2t})\notag\\
           &=\hat{F_2}(A_{2t},T_t) \tilde{F_2}(K_{2t},L_{2t}, E_{2t})
\end{align}
Note that climate capital is uniformly utilized in both sectors, whereas other factors and production technologies can be sector-specific.

 In a competitive market, profit-maximizing firms in both sectors should equate their marginal products to their prices:
\begin{align}
p_{1t}F_{1lt} &= p_{2t}F_{2lt} = w_t\\
p_{1t}F_{1kt} &= p_{2t}F_{2kt} = r_t \notag\\
p_{1t}F_{1Et} &= p_{2t}F_{2Et} = p_{Et} \notag\
\end{align}
where $F_{ijt}$ represents the partial derivative of sector $i$ production function with respect to rival input $j\in\{K,L,E\}$, and $p_{Et}$ denotes the energy price. 

In addition, production in both sectors can be utilized for either consumption or investment such that:
\begin{gather}
\label{eq:allocation}
Y_{1t} = C_{1t}+I_{1t}\\
Y_{2t} = C_{2t}+I_{2t} \notag\
\end{gather}
where $I_{it}$ is the output from sector $i$ used for investment.

Total nominal final output can thus be defined as:
\begin{align}
Y_t = p_{1t}Y_{1t} +  p_{2t}Y_{2t}\
\end{align}

\subsection{Investment production sector}
In the economy, there is also an intermediate investment sector that adopts a constant-to-scale technology and combines the production from two final sectors to produce final investment:
\begin{align}
\label{eq:invest}
I_t=F_I(A_{It},I_{1t}, I_{2t})
\end{align}
where $A_{It}$ denotes an exogenous investment-specific technical change to produce aggregate investment. We do not allow for a direct climate impact on investment production in Eq.(\ref{eq:invest}). However, temperature change indirectly influences investment because as it reduces final products supplied to produce investment\footnote{Allowing for the climate impact on investment explicitly will exacerbate the climate impact on growth rate. \citet{fankhauser2005climate} and \citet{dietz2015endogenous} show that such growth effect can be embodied in the climate impact on capital depreciation explicitly, analogous to modelling a climate impact on investment production. A climate impact on economic growth will acutely exacerbate economic losses compared to its frontier, whereas the level effect is more modest \citep{pindyck2021we,cai2023climate}. However, there still exist some intellectual gaps between modelling practices and empirical evidence to support the growth effect of climate change \citep{dell2012temperature,burke2015global,newell2021gdp}. Further, \citet{herrendorf2021structural} and \citet{garcia2021investment} find the exogenous investment-specific technical change plays a very limited role in driving long-run growth. Given all these, we only implicitly account for the climate impact on investment.}.

\subsection{Energy sector}
Using both capital $K_{Et}$ and labor $L_{Et}$, the intermediate energy sector produces with a constant-return-to-scale technology. Hence, energy production is given by:
\begin{align}
\label{eq:energy production}
E_t=A_{Et}\tilde{F_E}(K_{Et}, L_{Et})
\end{align}
Energy production is allocated to two final sectors as input. Following \citet{barrage2020optimal}, we assume that carbon-based energy can be unlimitedly supplied and therefore incurs zero Hotelling rents. Climate capital is in reality also used for energy production, but we make simplifications here considering that the energy sector accounts for a relatively small proportion in the economy.

Energy firms can choose to produce a fraction $\mu_t$ of energy with some zero-emission technologies at an additional abatement investment $\Theta_t(\mu_tE_t)$. For each unit of emission, firms are obliged to pay the carbon tax. Thus, the profits of energy producers are:
\begin{align}
\label{eq:profitenergy}
\Pi_t=p_{Et}E_{t}-[(1-\mu_t)E_t]\tau_{Et}-w_t L_{Et}-r_t K_{Et}-\Theta_t(\mu_tE_t)
\end{align}
where $\tau_{Et}$ denotes the carbon tax on uncontrolled carbon emissions from two final sectors $\{E_i^{unc}\}_{i=0}^t=\{(1-\mu_t)(E_{1i}+E_{2i})\}_{i=0}^t$. We only consider an aggregate carbon control rate $\mu_t$, and we discuss the implications of differentiated control rates in Section \ref{sec:Sensitivity and extensions}.

Again, in the competitive market, energy producers equate the marginal products of each input to their corresponding prices:
\begin{align}
(p_{Et}-\tau_{Et})F_{Elt}&=w_{t}\\
(p_{Et}-\tau_{Et})F_{Ekt}&=r_{t}\notag\
\end{align}

Moreover, the formulation in Eq.(\ref{eq:profitenergy}) implies that profit-maximizing energy producers are incentivised to equate the marginal benefit of avoided tax payment per unit of uncontrolled carbon emissions to marginal abatement costs:
\begin{align}
\tau_{Et}=\Theta_t^{'}(\mu_tE_t)
\end{align}

In each period, the productive factors are freely mobile across sectors:
\begin{align}
\label{eq:market clear}
L_{t} &= L_{1t}+L_{2t}+L_{Et}\\
K_{t} &= K_{1t}+K_{2t}+K_{Et}\notag\\
E_{t} &= E_{1t}+E_{2t}\notag\
\end{align}
where aggregate labor force $L_t$ is exogenously given at each period. 

\subsection{Carbon cycle and climate}
We follow the convention of climate economics literature in viewing temperature as a sufficient proxy for climate change. Specifically, atmospheric temperature change $T_t$ at period $t$ is determined by the historical path of carbon emissions after control $\{E_i^{unc}\}_{i=0}^t=\{(1-\mu_i)(E_{1i}+E_{2i})\}_{i=0}^t$, initial climate conditions $\bm{M}_0$ including atmospheric carbon concentrations, deep ocean temperatures, etc., and exogenous shifters $\{ \bm{\eta}_{i} \}_{i=0}^t$ such as land-based emissions:
\begin{equation}
  \label{eq:climate dynamics}
    T_{t}=\Phi(\bm{M}_0,E_0^{unc},E_1^{unc},...,E_i^{unc},\bm{\eta}_{0},...,\bm{\eta}_{t})
\end{equation}
where $\frac{\partial T_{t+j}}{\partial E_t^{unc}}$ holds for ${\forall} t,j \geq 0$. The very nature that climate change, proxied by $T_t$, hinges on the \emph{stock} of carbon emissions in the atmosphere hints that climate is a capital. Note that temperature is not exclusively dictated by current-period carbon emission, which is a \emph{flow} variable.

\subsection{Competitive equilibrium}
Now we can present the standard definition of competitive equilibrium in the economy, augmented with the climate system.
\begin{myDef}
A competitive equilibrium consists of a sequence of exogenously-given productivity $\{ A_{1t}, A_{2t}, A_{It}, A_{Et} \} _{t=0} ^ \infty$, a series of allocations $\{ C_{1t}, C_{2t}, I_{1t}, \\ I_{2t}, L_{1t}, L_{2t}, L_{Et}, K_{1,t+1}, K_{2,t+1}, K_{E,t+1}, E_{1t}, E_{2t}, \mu_t,T_t \} _{t=0} ^ \infty$, a set of prices $\{ r_{t}, \\ w_{t}, p_{1t}, p_{2t}, p_{Et}, \} _{t=0} ^ \infty$  and a series of policies $ \{\tau_{Et}\} _{t=0} ^ \infty$  such that in each period, given prices and policies:\\
(i) the household solves the utility-maximizing problem subject to the budget constraint,\\
(ii) firms in two final production sectors and two intermediate sectors (energy and investment) maximize profits,\\
(iii) temperature changes in line with the carbon cycle constraint, and\\
(iv) markets clear.
\end{myDef}

By virtue of the above definition, we now demonstrate what a first-best carbon tax is needed to decentralize the optimal allocation in the competitive equilibrium: 
\begin{proposition}
\label{pro:pro1}
The allocations $\{ C_{1t}, C_{2t}, I_{1t}, I_{2t}, L_{1t}, L_{2t}, L_{Et}, K_{1,t+1},K_{2,t+1}, \\K_{E,t+1}, E_{1t}, E_{2t}, \mu_t,T_t \} _{t=0} ^ \infty$, along with initial capital stock $K_0$, initial carbon concentrations $\bm{M}_0$, and climate shifters $\{ \bm{\eta}_{i} \}_{i=0}^t$ in a competitive equilibrium satisfy:
\begin{gather}
\label{eq:production 1 allocation}
F_1(A_{1t},T_t,L_{1t},K_{1t},E_{1t}) \geq C_{1t}+I_{1t}\\
F_2(A_{2t},T_t,L_{2t},K_{2t},E_{2t}) \geq C_{2t}+I_{2t}\\
F_C(C_{1t},C_{2t}) \geq C_{t}\\
\label{eq:capital accumulation}
F_I(A_{It},I_{1t},I_{2t})+(1-\delta)K_t \geq K_{t+1}+\Theta_{t}(\mu_t E_{t})\\
E_t \leq A_{Et}F_E(K_{Et},L_{Et})\\
T_t \geq \Phi(\bm{M}_0,(1-\mu_0)(E_{10}+E_{20}), ...,(1-\mu_t) (E_{1t}+E_{2t}), \bm{\eta}_{0},...,\bm{\eta}_{t})\\
L_{t} \geq L_{1t}+L_{2t}+L_{Et}\\
K_{t} \geq K_{1t}+K_{2t}+K_{Et}\\
\label{eq:energy allocation}
E_{t} \geq E_{1t}+E_{2t}
\end{gather}
Therefore, given an allocation that maximizes the household's net present utility Eq.(\ref{eq:utility}) and simultaneously satisfies constraints Eq.(\ref{eq:production 1 allocation})-Eq.(\ref{eq:energy allocation}), letting $\lambda_{It}$ the Lagrange multiplier on the capital accumulation constraint Eq.(\ref{eq:capital accumulation}), one can formulate a carbon tax equal to:
\begin{align}
\label{eq:carbontax}
\begin{split}
\tau_{Et}&=\underbrace{\Theta^{'}_t}_{Marginal \quad abatement \quad cost}\\ &=\underbrace{F_{I1t}F_{1Et}-\frac{F_{I1t}F_{1lt}}{F_{Elt}}=F_{I2t}F_{2Et}-\frac{F_{I2t}F_{2lt}}{F_{Elt}}}_{Marginal \quad product \quad of \quad energy}\\
&=\overbrace{(-1)\sum_{j=0}^\infty \beta^j \left( \underbrace {\frac{U_{C,t+j}}{\lambda_{It}} \frac{\partial C_{t+j}}{\partial C_{1,t+j}} \frac{\partial Y_{1,t+j}}{\partial T_{t+j}} \frac{\partial T_{t+j}}{\partial E_t^{unc}}}_{Impacts \quad on\quad sector \quad 1}+ \underbrace {\frac{U_{C,t+j}}{\lambda_{It}} \frac{\partial C_{t+j}}{\partial C_{2,t+j}} \frac{\partial Y_{2,t+j}}{\partial T_{t+j}}\frac{\partial T_{t+j}}{\partial E_t^{unc}}}_{Impacts \quad on\quad sector \quad 2} \right)}^{Marginal \quad benefit \quad of \quad carbon \quad abatement}     
\end{split}
\end{align}
such that the competitive market can achieve the optimal allocation as in the social planner problem, and the discounting of output damages in period $t$ is governed by:
\begin{align}
\frac{\lambda_{It}}{\beta\lambda_{I,t+1}}
&=F_{Ig,t+1}F_{gk,t+1} +(1-\delta)=F_{Is,t+1}F_{sk,t+1} +(1-\delta)
\end{align}
\end{proposition}

Proof: See \ref{sec:appendix:one}. Eq.(\ref{eq:carbontax}) describes the classic wisdom \citep{baumol1972taxation, golosov2014optimal} to address climate (environmental) externality. The optimal allocation requires \textbf{equating} marginal product of carbon-based energy \textbf{to} marginal abatement costs, \textbf{and also to} current-value marginal benefit of carbon abatement, namely climate impact due to an additional tone of carbon emission times minus one. To decentralize such optimal allocation, the global planner needs to levy a Pigouvian tax equal to $\Theta^{'}_t$.

However, the theoretical finding deviates from previous studies primarily in establishing investment, rather than consumption, as the core metric in the trade-off. Marginal abatement cost is by construction denominated by investment as previously introduced. The marginal product of energy is represented by how much final investment per unit of energy can produce net of the opportunity cost in energy production. The marginal impact of carbon emission is also denominated in terms of investment. Under optimal allocation, Pigouvian tax is numerically equal to the social cost of carbon. Social cost of carbon is conventionally defined as the shadow price of carbon measured by the utility of per unit of consumption \citep{nordhaus2014estimates,golosov2014optimal,barrage2020optimal}\footnote{Consumption-equivalent social cost of carbon can be generalized as $SCC_{CE}=\frac{\partial U}{\partial E_t}/\frac{\partial U}{\partial C_t}$, while investment-equivalent social cost of carbon is formulated as $SCC_{IE}=\frac{\partial U}{\partial E_t}/\frac{\partial U}{\partial I_t}$.}.  Thus, the common definition reflects the amount of consumption one would like to sacrifice today in order to reduce an additional unit of welfare-reducing emissions. 

Motivated by this result, we provide an alternative definition for social cost of carbon, \emph{i.e.} investment loss due to an additional tone of carbon emission. Recall in Section \ref{sec:Climate capital} that climate is a capital and carbon is an investment. If economic production is not utilized for the investment in abatement technology, it can be leveraged to accumulate physical capital. However, if no abatement technology is adopted, carbon will be overinvested. Eventually, the global mean temperature rises, generating an impact equivalent to a loss in physical capital. The pigouvian tax in Eq.(\ref{eq:carbontax}) seeks an optimal investment bundle that guarantees the equalized marginal product of each kind of investment. Therefore, the real cost of carbon is on investment rather than consumption, and climate change is due to disinvestment instead of malconsumption.

In fact, two definitions of social cost of carbon deviate only 
when investment and consumption are produced in different ways. Denoting the marginal value of investment $\lambda_{It}$ and that of consumption $\lambda_{Ct}$, one can establish the following identity: 
\begin{align}
\label{eq:MTR_CI}
\frac{\lambda_{It}}{\lambda_{Ct}}= \frac{F_{C1t}}{F_{I1t}}= \frac{F_{C2t}}{F_{I2t}}
\end{align}
The marginal value of investment relative to that of consumption is governed by how each final product is transformed into between consumption and investment. In one-sector climate-economy models without any other distortions like \citet{nordhaus2014estimates} and \citet{golosov2014optimal}, a unit of final production can be either consumed or invested without any differences. Thus, both $F_{C1t}/F_{I1t}$ and $F_{C2t}/F_{I2t}$ are cancelled out, and two definitions of social cost of carbon are numerically identical. 

Although social cost of carbon as a price is different, optimal allocation is fixed regardless of any numeraire. In addition, both definitions have their strengths and weaknesses. Consumption-equivalent social cost of carbon does not reflect the very nature that climate change is a drag on investment, but has convenient welfare implications because it demonstrates the consumption loss due to an additional tone of carbon emission. In comparison, investment-equivalent social cost of carbon cannot provide direct welfare interpretations. After all, people care consumption rather than investment\footnote{Having said that, investment-equivalent social cost of carbon can be utilized to calculate genuine saving (or comprehensive wealth) for measuring sustainable development \citep{arrow2004we,10.2307/26265518}.}.

\subsection{Drivers of structural change}
We induce structural change within both consumption and investment through the price effect\footnote{Existing studies generalize two broad forces behind structural change\textemdash the income effect and the price effect. The income effect captures that as income increases, so will the value-added share of products with a higher income elasticity. }. The price effect reflects that the value-added share of a certain product can also increase when its relative price goes up. We assume two final production sectors are only different in technological growth rate and climate vulnerability. Following \cite{herrendorf2014growth}, the production functions in both sectors adopt the Cobb-Douglas form with identical factor intensity\footnote{The price effect can also occur when sectors differ in capital intensity or the substitution elasticity between reproducible factors \citep{acemoglu2008capital,alvarez2017capital}. We abstract from these two possibilities for two reasons. First, \citet{herrendorf2015sectoral} showed that sectors with different productivity growths alone can fit well the post-war structural change in the United States. Second, the Baumol's cost disease, which is the focus of this paper, is concerned with the technological structure behind each sector.}:
\begin{align}
    \label{eq:prod2}
    Y_{1t} &=\hat{F_1}(A_{1t},T_t) K_{1t}^{\alpha}L_{1t}^{1-\alpha-\nu}E_{1t}^{\nu}\notag\\
    Y_{2t} &=\hat{F_2}(A_{2t},T_t) K_{2t}^{\alpha}L_{2t}^{1-\alpha-\nu}E_{2t}^{\nu}
\end{align}
where how climate capital interacts with knowledge capital in the production functions remains undefined without loss of generality.

In addition, both consumption and investment functions are assumed to be the CES form:
\begin{align}
\label{eq:consumption}
C_t &= \left(\omega_c^{\frac{1}{\epsilon_c}}  C_{1t}^{\frac{\epsilon_c-1}{\epsilon_c}} + (1-\omega_c)^{\frac{1}{\epsilon_c}}  C_{2t}^{\frac{\epsilon_c-1}{\epsilon_c}} \right) ^ {\frac{\epsilon_c}{\epsilon_c-1}}\\
\label{eq:investment}
I_t &= A_{It} \left(\omega_I^{\frac{1}{\epsilon_I}}  I_{1t}^{\frac{\epsilon_I-1}{\epsilon_I}} + (1-\omega_I)^{\frac{1}{\epsilon_I}}  I_{2t}^{\frac{\epsilon_I-1}{\epsilon_I}} \right) ^ {\frac{\epsilon_I}{\epsilon_I-1}}
\end{align}
where $\omega_c$ and $\omega_I$ are the weights of product $1$ in consumption and investment. $\epsilon_c$ and $\epsilon_I$ are the substitution elasticities between two products in producing final consumption and investment, both of which are less than unit. Note that consumption and investment can be different in terms of weight of each product, substitution elasticity, and investment-specific technological progress.

\begin{proposition}
\label{pro:pro2}
Given the production function in Eq.(\ref{eq:prod2}), absent carbon tax, it is straightforward to show that the relative price between two products in competitive market is pinned down by:
\begin{align}
\label{eq:relative price 1}
\frac{p_{2t}}{p_{1t}}=\frac{\hat{F_1}(A_{1t},T_t)}{\hat{F_2}(A_{2t},T_t)}
\end{align}
such that given the consumption and investment production functions as Eq.(\ref{eq:consumption}) and Eq.(\ref{eq:investment}), the share of product $i$ will increase in tandem with its relative price. Thus, structural change takes place via the price effect. Moreover, two definitions of social cost of carbon are linked by the transformation ratio:
\begin{align}
\label{eq:ratio}
\frac{SCC_{CE}}{SCC_{IE}}=\underbrace{\frac{1}{A_{It}}}_{ISTC} \times \underbrace{\frac{\left[\omega_{c}\hat{F_1}(A_{1t},T_t)^{\epsilon_c-1}+(1-\omega_{c})\hat{F_2}(A_{2t},T_t)^{\epsilon_c-1}\right]^{\frac{1}{\epsilon_c-1}}}{\left[\omega_{I}\hat{F_1}(A_{1t},T_t)^{\epsilon_I-1}+(1-\omega_{I})\hat{F_2}(A_{2t},T_t)^{\epsilon_I-1}\right]^{\frac{1}{\epsilon_I-1}}}}_{SC \quad interacted \quad with \quad climate}
\end{align}
\end{proposition}

Proof: See \ref{sec:appendix:one}. Proposition \ref{pro:pro2} presents two relative prices that are related to the Baumol's cost disease and social cost of carbon, respectively. 

\textbf{The first} is the relative price between two final products in Eq.(\ref{eq:relative price 1}), jointly determined by technological productivity growth and climate vulnerability. We consider the relative price under three different cases, and discuss their relevance to the Baumol's cost disease. Note that the substitution elasticity between two final products are less than unit in both Eq.(\ref{eq:consumption}) and Eq.(\ref{eq:investment}). 

\textbf{Case 1: Only technological productivity affects the relative price.} In structural change literature, climate vulnerability is normally not included. Thus, following the common assumption that technology enters the production function in multiplicative form, Eq.(\ref{eq:relative price 1}) boils down to: 
\begin{align}
\frac{p_{2t}}{p_{1t}}=\frac{A_{1t}}{A_{2t}}
\end{align}
Assume that sector $1$ has a robust productivity growth, while sector $2$ is stagnant in growth (like services). Thus, the relative price of product 2 to product 1 will increase over time. As products in both sectors are necessary in producing consumption and investment, more and more productive resources will flow into the non-progressive sector because of its increasing relative price, and the share of this sector is expanding in the economy. In the long run, the aggregate productivity growth slows down due to an growing sector with slow productivity growth. Thus, \emph{the Baumol's cost disease} occurs, as studied in \citet{ngai2007structural} and \citet{herrendorf2021structural}.

\textbf{Case 2: Only climate vulnerability affects the relative price.} Climate vulnerability between two sectors, determined by climate capital, can also influence the relative price, leading up to \emph{the Baumol's climate disease}. Assume that two final sectors have fixed knowledge capital so that technological productivity is constant in both sectors. Thus, Eq.(\ref{eq:relative price 1}) can be rewritten to:
\begin{align}
\frac{p_{2t}}{p_{1t}}=\frac{\hat{F_1}(T_t)}{\hat{F_2}(T_t)}
\end{align}
where $\hat{F_i}(T_t)$ reflects the ability of sector $i$ to produce to uniform climate capital $T_t$. As explained in Section \ref{sec:Climate capital}, it can be perceived as the climate damage function, and governs heterogeneous climate vulnerability. Thus, a higher level of climate capital $T_t$ implies a high damage level, and hence a lower level of $\hat{F_i}(T_t)$.  As climate capital is nonrival and nonexcludable, climate capital is overestimated and a higher level of global mean temperature is attained. Assume that economic production in sector 2 is more reliant on climate capital. Thus, given a higher temperature uniform to both sectors, $\hat{F_2}(T_t)$ is increasingly lower than $\hat{F_1}(T_t)$. Thus, the relative price of product 2 is climbing gradually. Again, because both products are necessary and cannot be well substituted, the expenditure on product 2 will increase in accordance. Thus,  given an expanding sector in the economy more vulnerable to climate change, the aggregate economy will become increasingly vulnerable to climate change. In effect, a sector that is more vulnerable to climate change is technically a sector with slower productivity growth.

Existing climate-economy models \citep[e.g.:]{dice2023, golosov2014optimal, barrage2020optimal} usually treat the economy as a single final sector, and study the incentives to address climate change. In this paper, we show that because climate capital is nonrival, nonexcludable and unpriced, it leads to overinvested carbon. This is the first incentive for carbon abatement in Proposition \ref{pro:pro1}. We add here that the Baumol's climate disease is another important incentive to cope with climate change seriously after accounting for heterogeneous climate vulnerability. Long-run economic growth is hampered by climate change due to unpriced climate capital, and this adverse impact can be potentially aggravated by the Baumol's climate disease.

\textbf{Case 3: Both affect the relative price.} In reality, technological productivity and climate vulnerability jointly determine the relative price as in Eq.(\ref{eq:relative price 1}). Suppose that sector 2 has lower productivity growth and higher climate vulnerability, its relative price will be higher than both Case 1 and Case 2. Put differently, \emph{the joint effect of Baumol's cost and climate diseases} is even more acute. Thus, failure to pricing climate capital can drag down aggregate economic growth severely, and there is a stronger incentive to disinvest carbon. Suppose that sector 2 has high productivity growth and also high climate vulnerability. Under the plausible assumption that the impact of productivity growth outweighs climate vulnerability, the relative price of product 2 to product 1 may also increase, but to an extent less than absent climate change. Baumol's cost and climate diseases are ameliorated. Thus, the urgency to address climate change is reduced. This reasoning may be surprising at first sight, but can be valid. A sector with higher productivity growth implies a higher resilience to climate change. Although the sector may be more vulnerable to climate change, it can soon recover from climate damage by rapidly accumulating knowledge capital.

\textbf{The second price} bridges two definitions of social cost of carbon, and in fact reflects the relative price of investment to consumption. As shown by Eq.(\ref{eq:ratio}), the ratios depends on investment-specific technical change $A_{It}$, how climate capital interacts with knowledge capital in each final sector ${F_i}(A_{it},T_t)$, the weight of product 1 in producing investment and consumption  $\omega_{I}$ and $\omega_{c}$, and the substitution elasticity between two products in investment and consumption $\epsilon_I$ and $\epsilon_c$. In climate-economy models with one final production sector, because the productions of investment and consumption are not specified, all items are cancelled out\footnote{Section \ref{sec:Sensitivity and extensions} examines a simplified case where there are two final sectors producing investment and consumption, respectively.}. The consumption-equivalent social cost of carbon is the same as the investment-equivalent one. Our specifications of investment and consumption production are admittedly not comprehensive. But they serve as good examples for illustrating that only under very strict condition would one expect two definitions of social cost of carbon are numerically identical. The real cost of carbon is on investment, and but can be denominated in terms of consumption after transformation.

\section{Calibration}
\label{sec:Calibration}
\subsection{Sector division}
We divide the whole economy into goods and services, motivated by two reasons. It is important to note that sector division in this paper is for heuristic purpose, rather than for approximating the real world perfectly.

\textbf{Reason 1:} On the climate-to-economy side, the services sector appears to be less vulnerable to weather shocks than the goods sector. Following \citet{burke2015global} but focusing on sectoral impact, we investigate
\begin{align}
    \Delta Y_{it}^j = h(T_{ij}) + g(P_{ij}) + \mu_i + \zeta_t + \theta_i t + \theta_{i2} t^2+\epsilon_{it}
\end{align}
where $\Delta Y_{it}^j$ is the output growth rate of sector $j$ (either goods or services) in country $i$ at year $t$, 
$T$ annual average temperature, $P$ precipitation, $\mu_i$ country-specific constant terms, $\zeta_t$ year fixed effects and $\theta_i t + \theta_{i2} t^2$ flexible country-specific time trends.

\begin{figure}[htp]
\centering
\begin{threeparttable}
       \begin{minipage}[t]{0.49\textwidth}
        \centering
        \includegraphics[width=1\textwidth]{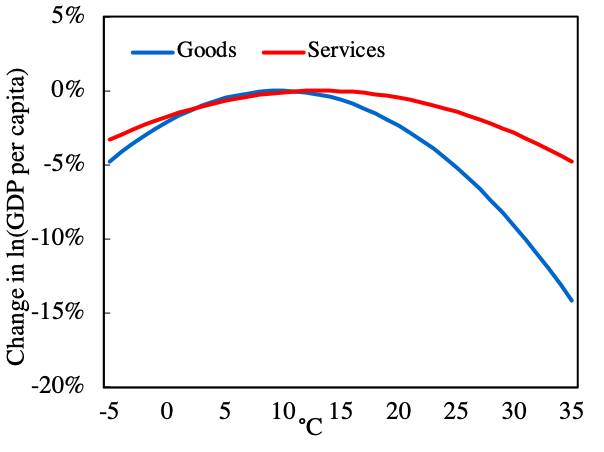}
        \centerline{\footnotesize{(a) Impact on goods and services}}
    \end{minipage}
    \begin{minipage}[t]{0.49\textwidth}
        \centering
        \includegraphics[width=1\textwidth]{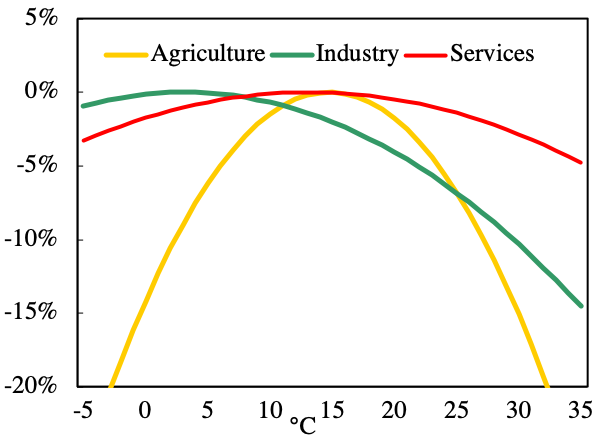}
        \centerline{\footnotesize{(b) Impact on three sectors}}
    \end{minipage}
    \begin{tablenotes}[flushleft]
    \footnotesize
    \item Notes: The figure shows how sectoral GDP growth can be affected by temperature shocks using the data from \citet{burke2015global}.         
    \end{tablenotes}
\end{threeparttable}
    \caption{Sectoral damages}\label{fig:damage}
\end{figure}

Figure \ref{fig:damage} shows when the global annual average temperature rises above around 13 degrees Celsius, the impact on the goods sector will soon be larger than on the services, and the gap is widened at a higher temperature (Panel a). Even when subtracting the agriculture sector, that is among the most vulnerable to temperature, weather shocks take a heavier toll on the left industry production than services (Panel b). Moreover, the adverse impact on industry can start at a temperature much lower than services. These results are consistent with \citet{casey2021understanding} and \citet{rudik_lyn_tan_ortiz-bobea_2021} that labor productivity can suffer a heavier loss under heat stress in outdoor activities  which are more concentrated in the goods sector.

\textbf{Reason 2:} The sluggish services sector may drive down aggregate productivity growth.

The services sector affects aggregate productivity growth. One-sector climate-economy models do not explicitly include this prospect. Evolving growth path in each sector combines to determine aggregate growth path \citep{foerster2022aggregate}, and hence an expanding sector with a lower-than-average growth rate is capable of slowing down aggregate growth, which is roughly the core of Baumol's cost disease \citep{baumol1967macroeconomics,nordhaus2008baumol}. In the past decades, a strand of literature has been actively engaged in leveraging the growth models to analyze the implications of rising services on long-run economic growth (e.g.: \citealp{duarte2010role,ngai2007structural,leon2020rise,duernecker2017structural}). However, the rise of services receives surprisingly deficient attention in the climate economics literature.

Given such division, we calibrate the model.

\subsection{Households}
The representative household maximize the population-weighted lifetime utility:
\begin{align}
\sum_{t=0}^\infty \beta^t N_t U(c_{1t},c_{2t})
\end{align}
where $N_t$ is aggregate population projections following the DICE model \citep{nordhaus2017revisiting}. $c_{1t}$ represents consumption of goods, and $c_{1t}$ consumption of services.

We use the literature to calibrate the consumption production function Eq.(\ref{eq:consumption}) (see Table \ref{tab:utility}). The gap in substitution elasticity between earlier and recent literature can be reconciled by two different accounting approaches matching the data, final expenditure and value added \citep{herrendorf2013two}. We take a benchmark value of 0.2 for parameter $\epsilon_C$ governing the substitution elasticity between goods and services in consumption. In addition, the services expenditure share in consumption is 0.8.  Overall, our chosen values are consistent with existing studies that observe the complementary relationship between goods and services in consumption as well as a relatively larger proportion of services.

\begin{table}[htb]
    \centering
        \caption{Parameters of the utility function\label{tab:utility}}
        \small
    \begin{tabular}{lcccc}
       \toprule
            Source & Calibrated to & $1-\omega_c$ & $\epsilon_c$ \\
       \midrule
            \citet{buera2009can} & USA & NA & 0.5 \\
            \citet{duarte2010role} & USA & 0.96 & 0.4 \\
           \citet{herrendorf2013two} & USA & 0.81 & 0.002\\
            \citet{moro2015structural} & USA & 0.95 & 0.4\\
            \citet{alder2022theory} & AUS, CAN, GBR \& USA & 0.60-0.77 & 0.00-0.17 \\ 
        \bottomrule
    \end{tabular}
\end{table}

\subsection{Production sectors}
We assume the Cobb-Douglas production technology for both final production sectors, with capital, labor and energy as inputs. Thus, the production functions are given as:
\begin{align}
\tilde{F}_1(L_{1t},K_{1t},E_{1t})=K_{1t}^\alpha L_{1t}^{1-\alpha-\beta}E_{1t}^\beta\\
\tilde{F}_2(L_{2t},K_{2t}, E_{2t})=K_{2t}^\alpha L_{2t}^{1-\alpha-\beta}E_{2t}^\beta\notag
\end{align}
The identical factor intensity in two sectors are assumed to simplify the model. If capital share differs in each sector, this would generate another force for structural change via the price effect, as discussed above\footnote{One could argue that energy share is higher in the goods sector as it is more energy-intensive. Given our main intention is to understand the interactions between climate change and structural change dynamics, we assume energy intensity is also equivalent in both sectors.}. Thus, following \citet{nordhaus2017revisiting} and \citet{golosov2014optimal}, we adopt a capital share of $\alpha=0.3$ and an energy share of $\beta=0.03$ for both sectors.

For the sectoral productivity growth and climate vulnerability, we assume:
\begin{align}
\hat{F_1}(A_{1t},T_t)=A_{1t}(1-D_1(T_t))\\
\hat{F_2}(A_{2t},T_t)=A_{2t}(1-D_2(T_t))\notag
\end{align}
where $D_i(T_t)$ represents the fraction of output loss due to temperature change $T_t$ in sector $i$ at time $t$. We follow both the DICE model \citep{nordhaus1992optimal,nordhaus2017revisiting} and \citet{golosov2014optimal} where climate damage enters the production function in multiplicative form. The departure from previous works is that we allow for differentiated impact of temperature rise on producing goods and services. To pin down the technological growth in each sector, we first abstract from climate damages. Then, we determine the relative ratio of the goods sector to the services from existing studies (see Table \ref{tab:produc}), and, taking this ratio as given, mimic the capital stock in the DICE model at 2100. This strategy guarantees the closest comparability to DICE at an identical wealth level. Let $\gamma_i$ denote the technological growth in sector $i$. Taking a benchmark value of 3 for $\gamma_1/\gamma_2$, we have $\gamma_1=10.86\%$ and $\gamma_2=3.62\%$.

\begin{table}[htb]
    \centering
    \begin{threeparttable}
        \caption{Parameters of sectoral productivity growth\label{tab:produc}}
        \small
    \begin{tabular}{lcccc}
       \toprule
            Source & Calibrated to & $\gamma_1/\gamma_2$ & $\gamma_a/\gamma_2$ & $\gamma_m/\gamma_2$\\
       \midrule
            \citet{leon2020rise} & USA & 5.43 & NA & NA \\
            \citet{comin2021structural} & USA & NA & 2.64 & 1.20 \\
            \citet{NBERw27731} & USA & NA & 4.17 & 1.75\\
        \bottomrule
    \end{tabular}
        \begin{tablenotes}[flushleft]
    \footnotesize
    \item Notes:  $\gamma_i$ denotes the technological growth in sector $i$, with 1 for goods, 2 for services, $a$ for agriculture and $m$ for manufacturing.      
    \end{tablenotes}
\end{threeparttable}
\end{table}

Now we move on to investment production function Eq.(\ref{eq:investment}). Recent studies delving into the structural change within investment arrive at consistent findings that exogenous investment-specific technical change has a minor or even negligible effect in long-run growth \citep{herrendorf2021structural,garcia2021investment,NBERw27731}. \citet{herrendorf2021structural} finds that the productivity growth in goods can be three times larger than investment-specific technical change. Therefore, in addition to assuming no climate impact, we further assume  for simplicity it remains constant, namely, $\gamma_{I,2015} = 0\% $. The investment-specific technology level in the initial period (in 2015) is normalized to one. Moreover, we adopt $\omega_I=0.57$ and $\epsilon_I=0.5$, both of which are consistent with existing studies (Table \ref{tab:invest}).

\begin{table}[htb]
    \centering
        \caption{Parameters of the final investment production function\label{tab:invest}}
        \small
    \begin{tabular}{lcccc}
       \toprule
            Source & Calibrated to & $\omega_I$ & $\epsilon_I$ \\
       \midrule
            \citet{herrendorf2021structural} & USA & 0.65 & 0.00 \\
            \citet{garcia2021investment} & Countries in PWT \& WIOD & 0.58 & 0.51 \\
            \citet{NBERw27731} & USA & 0.52 & 0.01\\ 
        \bottomrule
    \end{tabular}
\end{table}

For the energy sector, we also assume a Cobb-Douglas production function:
\begin{align}
E_t=A_{Et}K_{Et}^{\alpha_E} L_{Et}^{1-\alpha_E}
\end{align}
where the labor share is 0.403 adopted from \citet{barrage2020optimal}. In addition, the energy sector is assigned to the same labor-augmented productivity growth as the goods sector. 

For value added shares of goods and services in initial period, we refer to the World Development Indicators. Goods account for 33.57\% of aggregate value added, and services is 66.43\%. We further combine with first order conditions to determine the factor share in each sector in the initial period.

\subsection{Carbon cycle and climate models}
The current paper borrows the carbon cycle and climate model from the DICE model \citep{nordhaus2017revisiting}, which generates a warming of 3.1\celsius{} for a doubling of carbon concentrations in equilibrium and a transient climate sensitivity of 1.7\celsius.

First, the equations of the carbon cycle include three reservoirs (the atmosphere $M_t^{At}$, the upper oceans and the biosphere $M_t^{Up}$, and the deep oceans $M_t^{Lo}$):
\begin{equation}
    \begin{pmatrix} M_t^{At}\\M_t^{Up}\\M_t^{Lo} \end{pmatrix}=
    \begin{pmatrix} \phi_{11} & \phi_{21} & 0\\\phi_{12} & \phi_{22} & \phi_{32} \\ 0 & \phi_{23} & \phi_{33} \end{pmatrix}
    \begin{pmatrix} M_{t-1}^{At}\\M_{t-1}^{Up}\\M_{t-1}^{Lo}\end{pmatrix}+
    \begin{pmatrix} E_t^{M}+E_t^{Land}\\0\\0 \end{pmatrix}
\end{equation}
where $\phi_{ij}$ measures the carbon flow between reservoirs and $E_t^{Land}$ represents exogenous land carbon emissions.

Second, the increased carbon concentrations in the atmosphere elevate radiative forcing:
\begin{align}
\chi_{t} = \kappa \left[ \ln \left( M_t^{At}/ M_{1750}^{At}\right) / \ln(2) \right] + \chi_{t}^{Ex}
\end{align}
where $\chi_{t}^{Ex}$ is the exogenous forcing from other greenhouse gases.

Last, higher radiative forcing raises the atmospheric temperature $T_t$ and, indirectly, the deep ocean temperature $T_t^{Lo}$: 
\begin{equation}
    \begin{pmatrix} T_t\\T_t^{Lo} \end{pmatrix}=
    \begin{pmatrix} 1-\zeta_1\zeta_2-\zeta_1\zeta_3 & \zeta_1\zeta_3  \\ 1-\zeta_4 & \zeta_4 \end{pmatrix}
    \begin{pmatrix} T_{t-1}\\T_{t-1}^{Lo} \end{pmatrix}+
    \begin{pmatrix} \zeta_1\chi_t\\0 \end{pmatrix}
\end{equation}
where $\{\zeta_i\}_{i=1}^4$ are parameters governing heat exchange between the atmosphere and the ocean.

\subsection{Sectoral climate damages}
 In the DICE model, the damage function is calibrated to match a damage of 2.1\% damage of income at 3°C. We adopt the damage function as in DICE-2016, and suppose that the curvature of the damage function is identical in both sectors. Furthermore, our calibration matches that the relative climate damage in the goods sector is three times that of services and also that at a temperature rise of 3°C, the combined damage from two sectors amount to that of DICE.The damage function in the IAMs has always been one of the most uncertain components, and this concern also applies here. However, although the damage function is highly uncertain, our analysis only requires that sector is different in the productivity of climate capital. Therefore, the benchmark model adopts a damage function for output in each sector as follows:
\begin{align}
1-D_{1}(T_t)=\frac{1}{1+0.004352*T_t^2}\\
1-D_{2}(T_t)=\frac{1}{1+0.001414*T_t^2}\
\end{align}

\subsection{Abatement costs}
In the DICE model, the abatement cost function is directly governed by the control rate, whereas the abatement costs in this study are related to the unit of abated carbon emission. Thus, following \citet{barrage2020optimal}, we recalibrate the abatement cost function through a logistic approximation to the abatement cost curve implied by \citet{nordhaus2017revisiting}:

\begin{align}
\Theta_t(\mu_t E_t)=\frac{\overline{a}P_t^{backstop}}{1+a_t \exp(b_{0t}-b_{1t}(\mu_t E_t)^{b2})} \cdot (\mu_t E_t)\
\end{align}
where $P_t^{backstop}$ is the price for backstop technology.

Finally, we have $\overline{a}=0.7464$, $a_t=0.6561+0.8881t$, $b_{0t}=7.864-1.4858t$,
$b_{1t}=1.6791-0.3157t$ and $b_2=0.4207$. It should be noted that the abatement costs is denominated by investment product whereas the abatement costs in DICE adopt the final output as numeraire. We perform sensitivity tests for abatement costs, and find that the key insights in this study are not affected.

Further details on calibration are provided in \ref{sec:appendix:two}.

\section{Quantitative results}
\label{sec:Quantitative results}
In this section, we present the numerical results of the constructed climate-economy model. Two standard scenarios are considered throughout, a business-as-usual (BAU) scenario where the climate externality is completely ignored so that no abatement policy is adopted, and an optimal scenario where a carbon tax is adopted equating marginal abatement cost to marginal climate damage. We also consider, where necessary and for ease of exposition, an additional scenario with no climate (pure economic growth model) to demonstrate the economic frontier. We first present the damage level to explore whether the Baumol's climate disease is consistent with the theoretical results in Section \ref{sec:Model and theory}. Then, we quantify the social cost of carbon with two different definitions, namely, investment-equivalent and consumption-equivalent, and explain their implications.

\subsection{Climate damage and abatement incentive}

\begin{figure}[htp]
\centering
\begin{threeparttable}
    \begin{minipage}[t]{0.49\textwidth}
        \centering
        \includegraphics[width=1\textwidth]{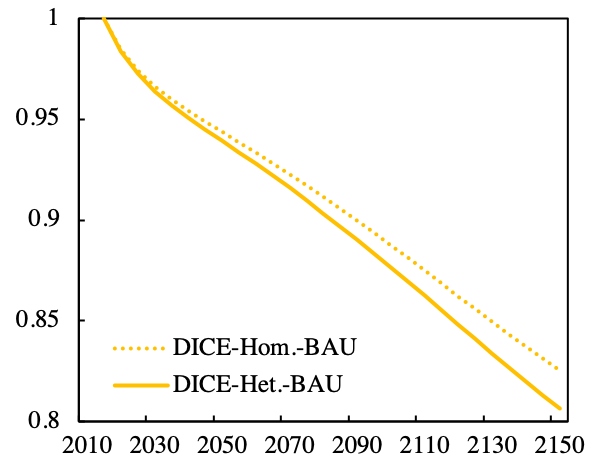}
        \centerline{\footnotesize{(a):Capital damage in DICE-like models}}
    \end{minipage}
    \begin{minipage}[t]{0.49\textwidth}
        \centering
        \includegraphics[width=1\textwidth]{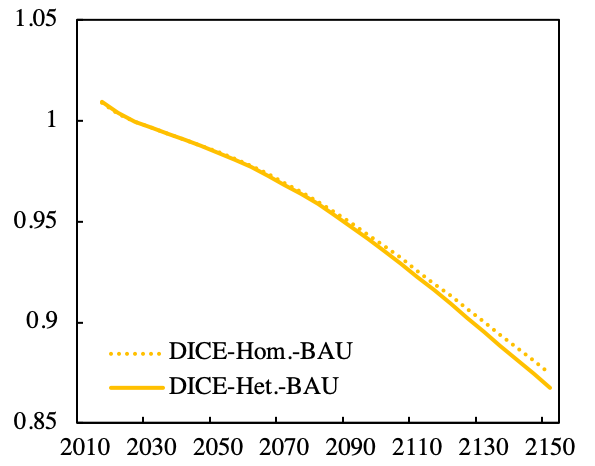}
        \centerline{\footnotesize{(b):Consumption damage in DICE-like models}}
    \end{minipage}
    \begin{minipage}[t]{0.49\textwidth}
        \centering
        \includegraphics[width=1\textwidth]{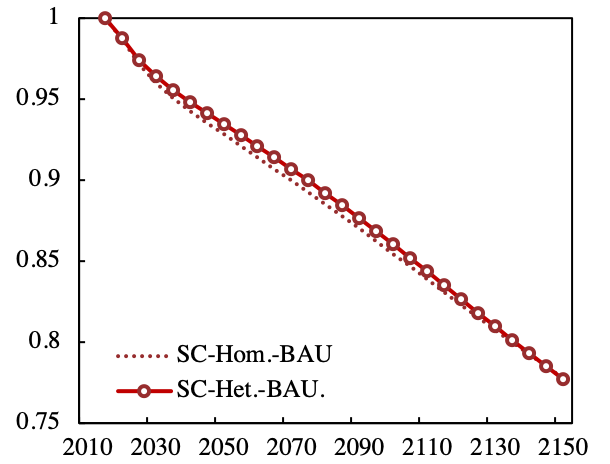}
        \centerline{\footnotesize{(c):Capital damage under  structural change}}
    \end{minipage}
    \begin{minipage}[t]{0.49\textwidth}
        \centering
        \includegraphics[width=1\textwidth]{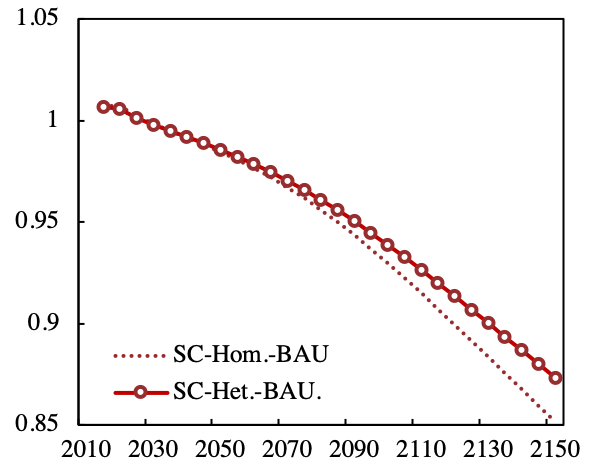}
        \centerline{\footnotesize{(d):Consumption damage under structural change}}
    \end{minipage}
    \begin{tablenotes}[flushleft]
    \footnotesize
    \item Notes: Panels (a)-(d) compare the results under both homogeneous(Hom.) and heterogeneous(Het.) climate vulnerability. The vertical axis describes the relative level of each economic variable in the business-as-usual scenarios compared to the their frontiers under no climate externality in each period. DICE means two identical sectors, SC means structural change.      
    \end{tablenotes}
\end{threeparttable}
    \caption{Capital and consumption damage}
    \label{Fig:damage}
\end{figure} 

Figure \ref{Fig:damage} displays the relative level of capital stock and consumption in the BAU scenario compared to their economic frontier without climate damage. DICE-like models denotes that two  sectors have identical productivity growth, while 'under structural change' means that the productivity growth rate in goods is three times that in services.

When two sectors have the same productivity growth, only the Baumol's climate disease is operating, as the Case 2 in Section \ref{sec:Model and theory}. Panels (a) and (b) show that climate damage is indeed higher under heterogeneous climate vulnerability, validating the impact of the Baumol's climate disease. When no abatement policy is adopted, capital stock is 5.62\% (6.07\%) lower in 2050, 11.25\% (12.34\%) lower in 2100, and 17.47\% (19.37\%) lower in 2150 than its frontier under homogeneous (heterogeneous) damage (Panel a). Aggregate consumption under heterogeneous damage is also in general lower than homogeneous one (Panel b), but the gap is narrower than capital damage. For DICE-like models,  the consumption level is 0.03 percentage points higher in 2050 when sectors are equally vulnerable, and this difference increases to 0.71
percentage points in 2150. 

If we assume the goods sector has a productivity growth two times faster than services, a different pattern is obtained. Under heterogeneous climate vulnerability, both capital and consumption damages are lower before the end of this century, compared to under homogeneous climate vulnerability. As investment consists of an increasing proportion of services, relative capital damage curve under heterogeneous damage overlaps with or is even slightly above that under homogeneous damage in initial periods (Panel c). Put differently, the trend that services is gaining its importance in investment makes economic growth more resilient. By comparison, aggregate consumption is elevated in the displayed period (Panel d), due to higher dependence on services in consumption than investment, and also to that capital damage is not aggravated under heterogeneous climate vulnerability in the displayed periods.

\begin{figure}[htp]
\centering
\begin{threeparttable}
    \begin{minipage}[t]{0.49\textwidth}
        \centering
        \includegraphics[width=1\textwidth]{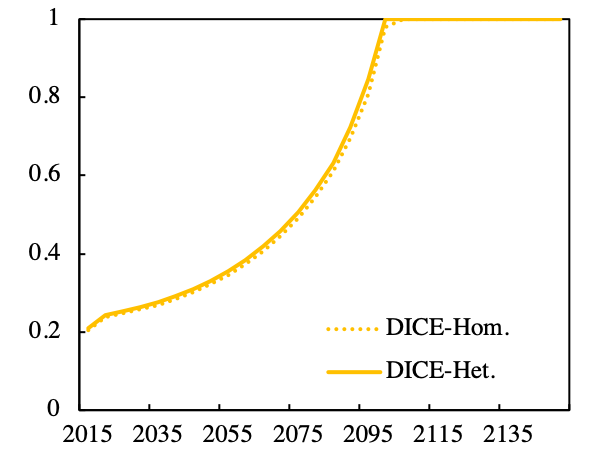}
        \centerline{\footnotesize{(a):DICE}}
    \end{minipage}
    \begin{minipage}[t]{0.49\textwidth}
        \centering
        \includegraphics[width=1\textwidth]{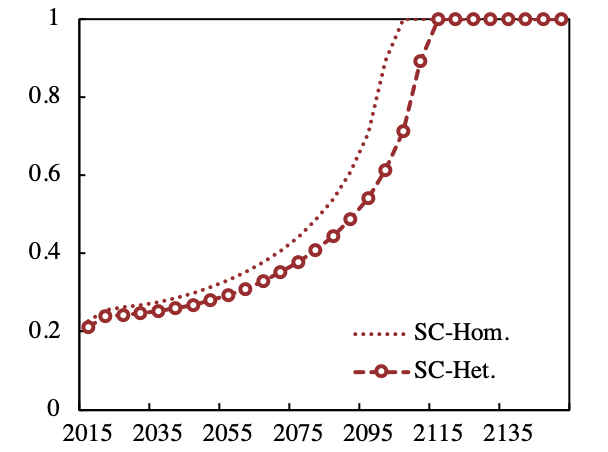}
        \centerline{\footnotesize{(b):Structural change}}
    \end{minipage}
    \begin{tablenotes}[flushleft]
    \footnotesize
    \item Notes: Panels (a)-(b) show the optimal abatement rate under both homogeneous(Hom.) and heterogeneous(Het.) climate vulnerability.      
    \end{tablenotes}
\end{threeparttable}
    \caption{Optimal abatement rate}
    \label{Fig:abatement}
\end{figure} 

Figure \ref{Fig:abatement} on optimal abatement rate confirms that the incentives to reduce investment into climate capital is influenced by the Baumol's cost and climate diseases. When only the  Baumol's climate disease is operating, the abatement curve is higher under heterogeneous climate vulnerability, although the its gap with under homogeneous climate vulnerability is not obvious (Panel a). But the time to achieve net zero emissions is advanced by five years. In comparison, if accounting for rising services, the abatement policy is less strict under heterogeneous climate vulnerability, as reflected in the lower circle-dash line (Panel b). It is suggested that the social planner should put off the time to achieve net zero emissions by 10 years. The reason is that although the goods sector is more vulnerable to climate change, its productivity growth rate is also higher than services. Thus, climate change ameliorates the impact of the Baumol's cost disease.

\subsection{Counterfactual experiments}
Previous numerical exercises are based on the observations that the goods sector is more climate-vulnerable than the services sector, whereas we analyze in Case 3 that the climate impact can be either amplified or ameliorated by the relative price. Because this paper cannot exhaust all sector divisions, it considers a counterfactual case assuming that the climate impact on the services sector is three times that on goods. In this case, the relative price between two sectors will be further increased.

\begin{figure}[htp]
\centering
\begin{threeparttable}
    \begin{minipage}[t]{0.49\textwidth}
        \centering
        \includegraphics[width=1\textwidth]{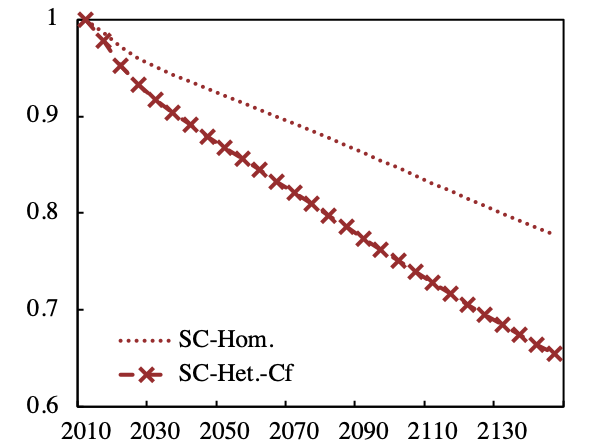}
        \centerline{\footnotesize{(a):Capital damage}}
    \end{minipage}
    \begin{minipage}[t]{0.49\textwidth}
        \centering
        \includegraphics[width=1\textwidth]{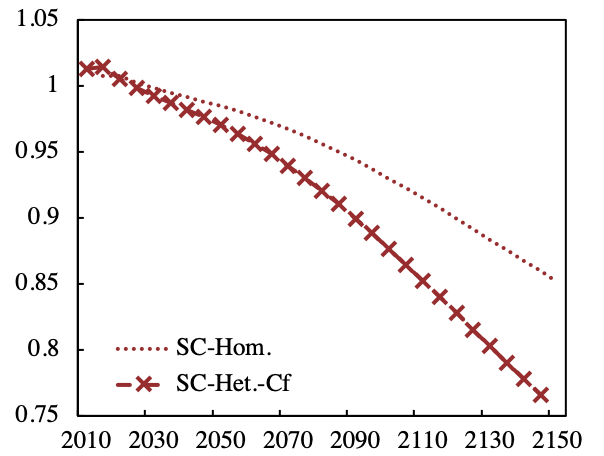}
        \centerline{\footnotesize{(b):Consumption damage}}
    \end{minipage}
    \begin{tablenotes}[flushleft]
    \footnotesize
    \item Notes: Panels (a)-(b) compare the results under both homogeneous(Hom.) and heterogeneous(Het.) climate vulnerability. The vertical axis describes the relative level of each economic variable in the business-as-usual scenarios compared to the their frontiers under no climate externality in each period. The services sector is assumed to be, counterfactually, more climate-vulnerable than the goods sector. 
    \end{tablenotes}
\end{threeparttable}
    \caption{Counterfactual capital and consumption damages}
    \label{Fig:damage2}
\end{figure} 

Figure \ref{Fig:damage2} displays that both capital and consumption are considerably reduced when heterogeneous climate vulnerability aggravates the Baumol's cost disease. Climate change will decrease capital stock by 14.55\% in 2100 if its impact is uniform to each sector. By comparison, the capital loss will additionally increase by 9.20 percentage points to 23.75\% under heterogeneous climate vulnerability. The impact on capital stock brings about a further drop of 4.14 percentage points in consumption by the end of this century. As a result, the optimal strategy to cope with climate change is to achieve net-zero emissions in 2085, twenty years ahead of the time under homogeneous vulnerability.

\subsection{Social cost of carbon: consumption equivalent v.s. investment equivalent}
 
\begin{figure}[htp]
\centering
\begin{threeparttable}
        \begin{minipage}[t]{0.49\textwidth}
        \centering
        \includegraphics[width=1\textwidth]{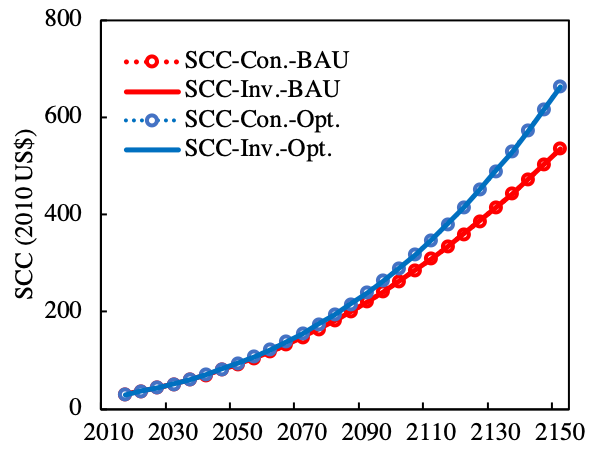}
        \centerline{\footnotesize{(a) DICE-like models}}
    \end{minipage}
    \begin{minipage}[t]{0.49\textwidth}
        \centering
        \includegraphics[width=1\textwidth]{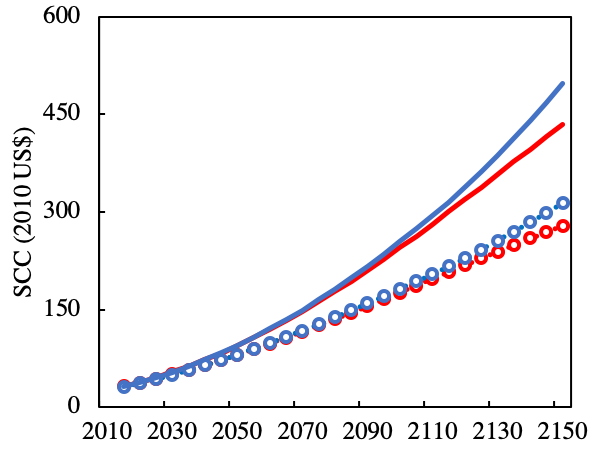}
        \centerline{\footnotesize{(b) Structual change}}
    \end{minipage}
    \begin{tablenotes}[flushleft]
            \footnotesize
    \item Notes: Two definitions of social cost of carbon are presented\textemdash consumption equivalent social cost of carbon (SCC-Con.) and investment equivalent one (SCC-Inv.). BAU denotes the business-as-usual scenario (no abatement), and Opt. denotes the optimal scenario.
    \end{tablenotes}
\end{threeparttable}
    \caption{Social cost of carbon}
    \label{fig:SCC}
\end{figure}

 In Section 4, we theoretically show that the actual trade-off between abatement cost and climate damage resides in how the final investment is affected, and propose investment-equivalent social cost of carbon in addition to consumption-equivalent one. In this part, we quantify the social cost of carbon under both definitions (Figure \ref{fig:SCC}). For better illustration, we compare DICE-like models with models under structural change.

Two definitions of social cost of carbon generate identical numerical values in the DICE-like models (Panel a). In a symmetric world, a unit of final production can be used to produce the same amount of investment or consumption, and therefore one can arbitrarily choose either definition. In contrast, under structural change (Panel b), there is an obvious divergence. In 2100, the social cost of carbon  in the optimal scenario comes to a investment loss of \$254/tCO\textsubscript{2}, or equivalently a consumption loss of \$182/CO\textsubscript{2}. In 2150, this gap further expands, with investment-equivalent social cost of carbon climbing to \$497/tCO\textsubscript{2}, 59\% higher than consumption-equivalent one. Because two final production sectors differ in productivity growth rate, climate vulnerability and also their compositions in consumption and investment, one unit of final production can no longer be converted to an equalized amount between investment and consumption. More concretely, since the production of consumption is assumed to be more heavily reliant on services, which has a lower productivity growth, its price relative to investment will rise, eventually leading to a lower value of social cost of carbon denominated by consumption.

\section{Sensitivity and extensions}
\label{sec:Sensitivity and extensions}
In this section,  we first test the sensitivity of our results by changing the values of some key parameters. Then, we extend with two alternative models.

\subsection{Sensitivity analyses}

\begin{figure}[htp]
\centering
\begin{threeparttable}
     \begin{minipage}[t]{0.49\textwidth}
        \centering
        \includegraphics[width=1\textwidth]{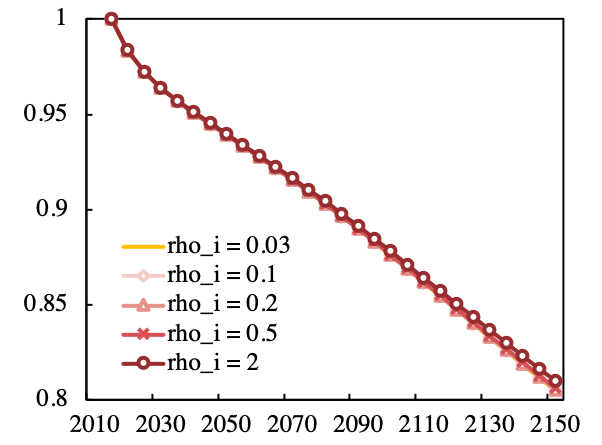}
        \centerline{\footnotesize{(a) DICE-like models}}
    \end{minipage}
    \begin{minipage}[t]{0.49\textwidth}
        \centering
        \includegraphics[width=1\textwidth]{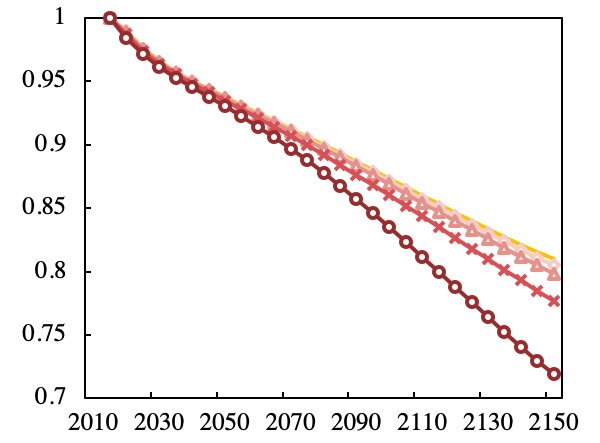}
        \centerline{\footnotesize{(b) Structural change}}
    \end{minipage}
    \begin{tablenotes}[flushleft]
        \footnotesize
    \item Notes: The vertical axis shows the relative level of capital stock under heterogeneous damage to its economic frontier (without climate damage). rho\_i  is the substitution elasticity between goods and services in producing final investment.       
    \end{tablenotes}
\end{threeparttable}
    \caption{Capital damage under different substitution elasticities}
    \label{fig:elasticity}
\end{figure}

First, the substitution elasticity between two sectors can affect the net impact of the Baumol's cost disease. We focus on the elasticity in investment production. In general, extant studies are consistent that the substitution elasticity between goods and services is low. In our baseline model, we choose a value of 0.5. Some studies argue that this value could be even lower and close to zero. Thus, we test the robustness of our results by choosing a value of 0.03, 0.1 and 0.5. Also, we perform the model with a value of 2 as a comparison, where two products can easily substitute each other in formulating the final investment. Results are displayed in Figure \ref{fig:elasticity}. Changing the substitution elasticity can work through altering both the price effect and the potential to transit to a less vulnerable sector consisting of more services.

When sectors are only different in climate vulnerability, a lower-than-unit substitution elasticity yields a lower level of capital stock, although changes are negligible (Panel a). Notably, when goods and services are assumed to be substitutes (rho\_i as 2), we observe that relative climate damages are slightly alleviated in the DICE-like models. Once accounting for the rising services in structural change model, we find that a lower elasticity is associated with a lower relative damage. Because a lower elasticity can generate a more severe Baumol's cost disease, the climate impact is further alleviated by the Baumol's cost disease. Moreover, when two products are considered as substitutes, the relative damage level is considerably aggravated. This should not be surprising because when two products can well substitute each other, the Baumol's cost disease is no longer operating and hence the economy is more prosperous absent climate impact. Higher production brings about more carbon emissions and consequently further increases global mean temperature, giving rise to higher damage levels.

\begin{figure}[htp]
\centering
\begin{threeparttable}
        \begin{minipage}[t]{0.49\textwidth}
        \centering
        \includegraphics[width=1\textwidth]{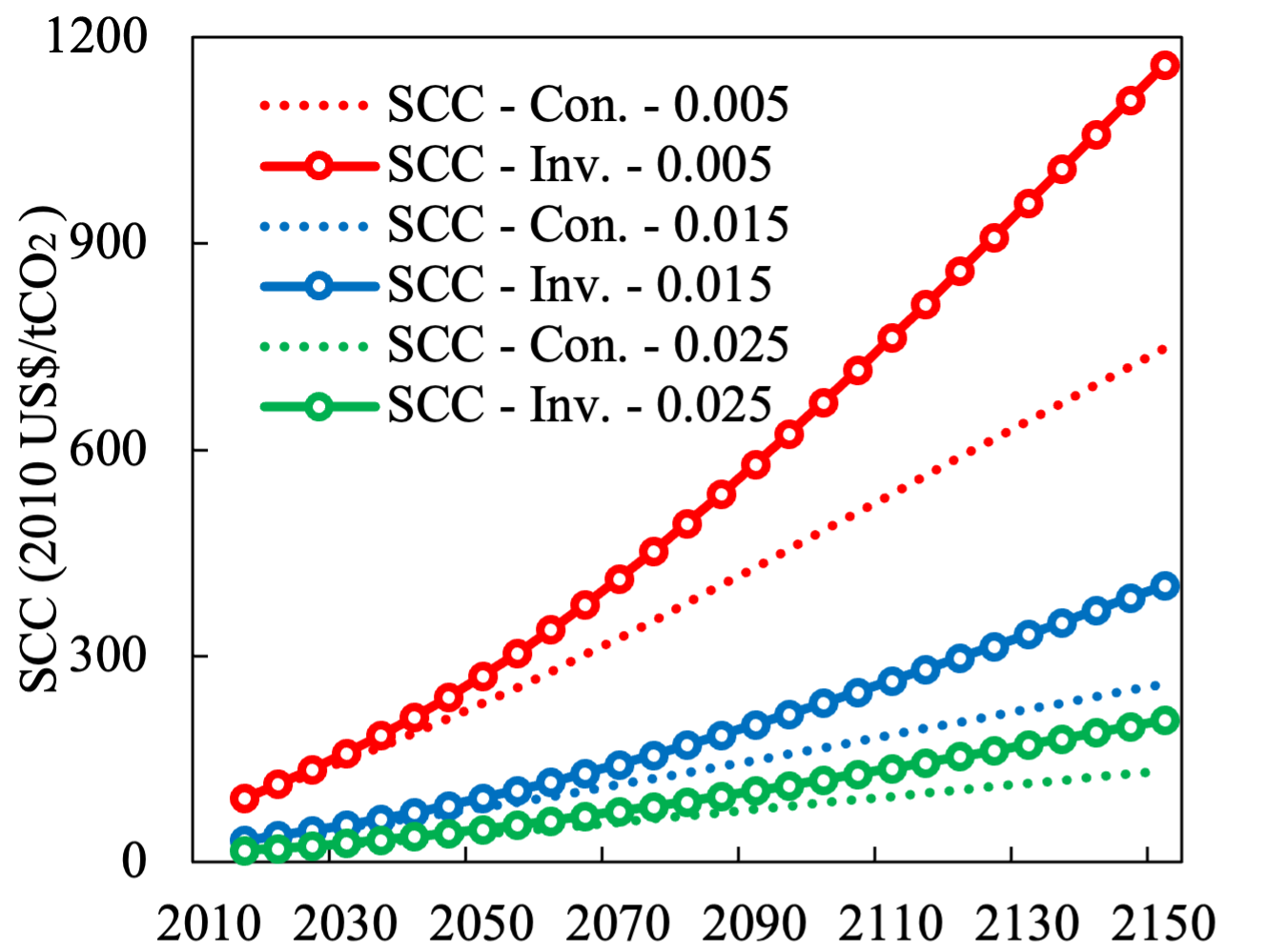}
        \centerline{\footnotesize{(a) Business-as-usual scenario}}
    \end{minipage}
    \begin{minipage}[t]{0.49\textwidth}
        \centering
        \includegraphics[width=1\textwidth]{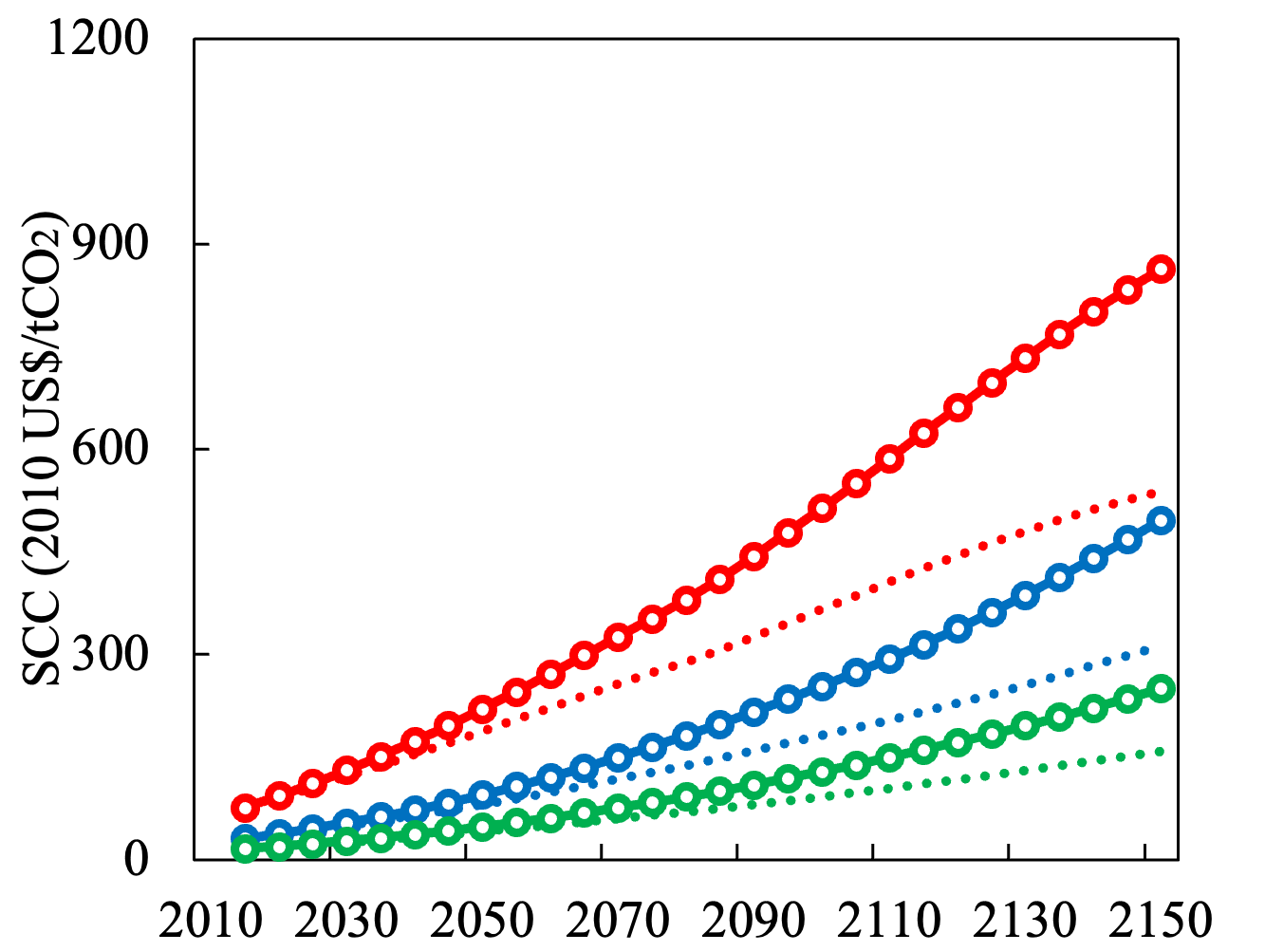}
        \centerline{\footnotesize{(b) Optimal scenario}}
    \end{minipage}
    \begin{tablenotes}[flushleft]
            \footnotesize
    \item Notes: Two definitions of social cost of carbon are presented in the figure, which are consumption equivalent social cost of carbon (SCC-Con.) and investment equivalent one (SCC-Inv.). Results are reported choosing different values of social time preference (0.005, 0.015 or 0.025).
    \end{tablenotes}
\end{threeparttable}
    \caption{Social cost of carbon and the discount rate}
\end{figure} 

Second, the social time preference rate is a key factor in driving the results of the social cost of carbon and debates around it abound \citep{barrage2018careful}. We are specifically interested in whether such choices will also matter to the gap between two definitions of social cost of carbon. We address this problem by altering the benchmark value of the social time preference from 0.015 (which is used by the DICE model) to a percentage point higher and lower. Figure 15 shows that when a lower social time preference is chosen, the implied social cost of carbon under both scenarios will climb dramatically. Moreover, the gap is amplified between the consumption equivalent social cost of carbon and the investment equivalent one. For example, the gap in 2100 between two definitions is enlarged from \$34/tCO$_t$ (0.025) to \$52/tCO$_t$ (0.015) and further to \$187/tCO$_t$ (0.005) under the business-as-usual scenario. A similar picture is observed in the optimal scenario. 

\subsection{Alternative models}
We present two alternative models and discuss their implications. All proofs are included in the appendix.

\textbf{Case 1: Consumption and investment.} The model can be simplified to a two-sector version that produces consumption and investment respectively. The two-sector growth model resembles in structure that in \citet{greenwood1997long} and \citet{foerster2022aggregate}, and enables us to demonstrate that the deviation between the consumption-equivalent and investment-equivalent social costs of carbon originates from the detailed consideration of the distinct procedures for producing consumption and investment. It can be shown that the optimal carbon tax in this case is pinned down by:
\begin{align}
\Theta^{'} &=F_{IEt}-\frac{F_{Ilt}}{F_{Elt}}\notag\\
&=(-1)\sum_{j=0}^\infty \beta^j (\underbrace {\frac{\lambda_{C,t+j}}{\lambda_{It}}\frac{\partial Y_{C,t+j}}{\partial T_{t+j}}}_{Impact \quad on \quad consumption} + \underbrace{\frac{\lambda_{I,t+j}}{{\lambda_{It}}} \frac{\partial Y_{I,t+j}}{\partial T_{t+j}}}_{Impact \quad on \quad investment} ) \frac{\partial T_{t+j}}{\partial E_t^{unc}}
\end{align}

This identity also establishes the shadow price of investment as denomination, but the climate damage aggregates both consumption damage and investment damage. In the baseline model, energy input should be utilized to produce two final products first, and then two products are combined to produce the aggregate investment. By comparison, the benefits of one additional tone of carbon emission is now represented by how much aggregate investment it can straightly produce. To sum up, the trade-off still centers on investment instead of consumption.

\textbf{Case 2: Differential abatement costs between sectors.} In the model, we do not allow for the more realistic case that each production sector can be characterized by different abatement costs \citep{gillingham2018cost}. This can be easily included in the model by introducing two individual abatement cost functions. In so doing, one can establish that the marginal abatement cost in each sector should be equalized and also equal to marginal climate damages:
\begin{equation}
    \begin{split}
        \Theta_{gt}^{'}= \Theta_{st}^{'} =  MD
    \end{split}
\end{equation}
where $\Theta_{it}$ denotes the abatement cost function in sector $i$ and $MD$ the marginal climate damages aggregating both sectors. It is of practical interest to include differentiated abatement functions in each sector\footnote{\citet{vogt2018starting} further shows that after accounting for the value of abatement capital in the future, optimal marginal abatement cost can differ in each sector. We do not include this concern in the current paper, but it can be an interesting extension.}.

\section{Discussion and Conclusion}
 \label{sec:Conclusion}
This paper establishes a dynamic general equilibrium model with endogenous structural change. Using the model,  we investigate optimal carbon abatement under structural change.

Three reasons stand out for reducing carbon investment in the atmosphere. First, climate is an essential capital used in economic production, but it is unpriced, nonrival, and nonexcludable. Thus, climate capital is disinvested by carbon emissions, and too much carbon emissions imply a deteriorating productive base, which can be conceived as reduction in physical capital. For modelling practice, climate capital proxied by global mean temperature is characterized by decreasing returns to scale. Second, the heterogeneous productivity of climate capital amplifies the productivity difference between sectors, leading to Baumol's climate disease. Third, Baumol's climate disease can either alleviate or exacerbate Baumol's cost disease, depending on the combination of productivty growth and climate vulnerability. Whenever there is a sector with higher climate vulnerability and lower productivity growth, higher abatement is required. The first reason is well-established \citep{nordhaus1992optimal}. The second and third are new and can call for more stringent greenhouse gas emission reduction.

The model used is stylized for maximum transparency and insight. This implies that key factors that may well affect optimal climate policy are omitted from the analysis. The model has one region only. This not only reduces the public bad nature of carbon dioxide emissions, it also excludes any effects of international trade and investment. The model has only two sectors. This makes the model less suitable as a representation of less developed countries. Structural economic change is one of the drivers of the Environmental Kuznets Curve. The upward phase of the EKC is hard to capture without agriculture. Agriculture is, of course, also the sector most exposed to the impacts of climate change. Furthermore, the model abstracts from distortions of the capital market and prior tax distortions. Such distortions surely interact with climate policy and may well with Baumol's cost and climate diseases. Finally, technological progress is assumed to be exogenous. The innovation public good interacts with the climate public bad. And, as technological progress is at the heart of Baumol's cost disease, endogenizing technological progress may well affect the results presented here. All these are deferred to future research.

Until those complications are studied in detail, we conclude that climate change causes a phenomenon that is similar to Baumol's cost diseases. Baumol's climate disease, and its interaction with Baumol's cost diseases when two diseases are complementary, increase the negative impacts of climate change and so justifies more stringent climate policy.

\section*{Acknowledgement}
The authors are grateful to Amalavoyal Chari, Pawel Dziewulski, L.Alan Winters, Huiying Ye, and all participants at the EAERE annual conference 2023 for useful comments and feedback. We also thank Timo Boppart and Georg Duernecker for kindly sharing their knowledge used in calibrating the model. In addition, Fangzhi Wang and Hua Liao appreciate the funding from National Science Fund for Distinguished Young Scholars (No.71925008) and National Natural Science Foundation of China (No.72293603). The views expressed are our own and do not reflect the views of the supporting agencies or authors' affiliations.
\appendix

\section{Proofs}
\label{sec:appendix:one}

The first step is to show how capital accumulation equation Eq.(\ref{eq:capital accumulation}) is obtained. We start with the representative household's problem and associated first order conditions. We assume throughout the solution to the household's problem is interior.  

The representative household seeks to maximize lifetime utility according to:
\begin{align}
U_{0} \equiv \sum_{t=0}^{\infty} \beta^t U(C_{t})
\end{align}
The household faces the following two constraints Eq.(\ref{eq:consumption production}) and Eq.(\ref{eq:budget constraint}).

Eq.(\ref{eq:consumption production}) describes how each kind of product is tranformed into final aggregate consumption, and Equation Eq.(\ref{eq:budget constraint}) is the budget constraint faced by the household. Letting $\zeta_t$ and $\gamma_t$ be the Lagrange multiplier on Eq.(\ref{eq:consumption production}) and Eq.(\ref{eq:budget constraint}) at time $t$ respectively, the first order conditions are given by:

[$C_t$]:
\begin{align}
\beta^t U_{Ct}=\zeta_t
\end{align}

[$C_{1t}$]:
\begin{align}
\zeta_t F_{C1t}=\gamma_t p_{1t}
\end{align}

[$C_{2t}$]:
\begin{align}
\zeta_t F_{C2t}=\gamma_t p_{2t}
\end{align}

[$K_{t+1}$]:
\begin{align}
\gamma_{t+1} (1+ r_{t+1} -\delta) =\gamma_t
\end{align}

For the producers of goods and services, their problems to maximize profits are:
\begin{align}
\max_{K_{it},L_{it},E_{it}} p_{it}Y_{it}-r_{t}K_{it}-w_{t}L_{it}-p_{et}E_{it}, \quad where \quad i \in \{1,2\}
\end{align}
Thus, the first order conditions to each input factor satisfy:
\begin{gather}
\label{eq:factorprice}
p_{1t}F_{1lt} = p_{2t}F_{2lt} = w_t\\
p_{1t}F_{1kt} = p_{2t}F_{2kt} = r_t \notag\\
p_{1t}F_{1Et} = p_{2t}F_{2Et} = p_{Et} \notag\
\end{gather}

The investment producer solves:
\begin{align}
\label{eq:investment profits}
\max_{I_{1t},I_{2t}} p_{It}F_{I}(A_{It}, I_{1t}, I_{2t})-p_{1t}I_{1t}-p_{2t}I_{2t}
\end{align}
where the price of final investment product in each period is normalized to one. The corresponding first order conditions are:
\begin{gather}
F_{I1t} = p_{1t}\\
F_{I2t} = p_{2t}\notag
\end{gather}

For the energy sector, the representative firm solves:
\begin{align}
\label{eq:energy profits}
\max_{K_{Et}, L_{Et}, \mu} \Pi_t=p_{Et}A_{Et}F_E(K_{Et}, L_{Et})-[(1-\mu_t)E_t]\tau_{Et}-w_t L_{Et}-r_t K_{Et}-\Theta_t(\mu_tE_t)
\end{align}
The associated first-order conditions are:
\begin{gather}
(p_{Et}-\tau_{Et})F_{Elt}=w_{t}\\
(p_{Et}-\tau_{Et})F_{Ekt}=r_{t}\notag\\
\tau_{Et}=\Theta_t^{'}(\mu_tE_t)\notag\
\end{gather}

Assume that the government only levies carbon tax and makes a lump-sum transfer to the household:
\begin{align}
\label{eq:transfer}
Tran_t=[(1-\mu_t)E_t]\tau_{Et}
\end{align}

Adding up the household's budget constraint (\ref{eq:budget constraint}), the definition of energy profits (\ref{eq:energy profits}), and the government transfer (\ref{eq:transfer}), we have:
\begin{align}
p_{gt}C_{gt}+p_{st}C_{st}+K_{t+1} \leq w_t (L_t-L_{Et}) + (1-\delta)K_t+  p_{Et}E_t + r_t (K_t-K_{Et}) \notag\\ -\Theta_t(\mu_tE_t) 
\end{align}
Substituting into the market clearing conditions as per capital, labor and energy Eq.(\ref{eq:market clear}) gives:
\begin{align}
p_{gt}C_{gt}+p_{st}C_{st}+K_{t+1} \leq w_t (L_{gt}+L_{st}) + (1-\delta)K_t+  p_{Et}(E_{gt}+E_{st})\notag\\  +r_t (K_{gt}+K_{st})-\Theta_t(\mu_tE_t) 
\end{align}
Invoking the factor prices based on first order conditions (\ref{eq:factorprice}) yields:
\begin{align}
p_{gt}C_{gt}+p_{st}C_{st}+K_{t+1} \leq p_{gt}F_{glt}L_{gt}+p_{st}F_{slt}L_{st} + (1-\delta)K_t+ p_{gt}F_{gEt}E_{gt}\notag\\+p_{st}F_{sEt}E_{st}
 +p_{gt}F_{gkt}K_{gt}+p_{st}F_{skt}K_{st}-\Theta_t(\mu_tE_t) 
\end{align}
Substituting the Euler's theorem based on the assumption of constant returns to scale in two final production sectors gives:
\begin{align}
p_{gt}C_{gt}+p_{st}C_{st}+K_{t+1} \leq p_{gt}Y_{gt}+p_{st}Y_{st}+ (1-\delta)K_t+\Theta_t(\mu_tE_t) 
\end{align}
Recalling the utilization rule of final production (\ref{eq:allocation}), the above equation can be rewritten to:
\begin{align}
K_{t+1} \leq p_{gt}I_{gt}+p_{st}I_{st}+ (1-\delta)K_t+\Theta_t(\mu_tE_t) 
\end{align}
Revoking the Euler's theorem based on the assumption of constant returns to scale in the investment production sector:
\begin{align}
K_{t+1} \leq I_t+ (1-\delta)K_t+\Theta_t(\mu_tE_t) 
\end{align}
This gives the capital accumulation constraint as in (\ref{eq:capital accumulation}). The carbon cycle constraint, the consumption aggregation constraint, the investment producer's budget constraint and the energy producer's budget constraint all hold by definition in competitive equilibrium. 

Thus, the social planner problem can be established as:
\begin{equation}
\begin{split}
  &\max \sum_{t=0}^{\infty} \beta^t U(C_{t})\\
&+ \sum_{t=0}^{\infty} \beta^t \lambda_{1t} \left[ F_1(A_{1t},T_t,L_{1t},K_{1t},E_{1t}) - C_{1t}-I_{1t} \right]\\
&+ \sum_{t=0}^{\infty} \beta^t \lambda_{2t} \left[ F_2(A_{2t},T_t,L_{2t},K_{2t},E_{2t}) - C_{2t}-I_{2t} \right]\\
&+ \sum_{t=0}^{\infty} \beta^t \lambda_{Ct} \left[ F_C (C_{1t},C_{2t})-C_t\right]\\
&+ \sum_{t=0}^{\infty} \beta^t \lambda_{It} \left[ F_I(A_{It},I_{1t},I_{2t})+(1-\delta)K_t - K_{t+1}-\Theta_{t}(\mu_t E_{t}) \right]\\
&+ \sum_{t=0}^{\infty} \beta^t \xi_t \left\{ T_t - \Phi[\bm{M}_0,(1-\mu_0)(E_{10}+E_{20}), ...,(1-\mu_t) (E_{1t}+E_{2t}), \bm{\eta}_{0},...,\bm{\eta}_{t}]\right\}\\
&+ \sum_{t=0}^{\infty} \beta^t \chi_{lt} \left[ L_{t} - L_{1t}-L_{2t}-L_{Et}\right]\\
&+ \sum_{t=0}^{\infty} \beta^t \chi_{kt} \left[ K_{t} - K_{1t}-K_{2t}-K_{Et}\right]\\
&+ \sum_{t=0}^{\infty} \beta^t \chi_{Et} \left[ A_{Et}F_E(K_{Et},L_{Et}) - E_{1t} - E_{2t} \right]  
\end{split}
\end{equation}

The F.O.C w.r.t. $C_{1t}$ is:
\begin{align*}
\lambda_{Ct}\frac{\partial C_t}{\partial C_{1t}}=\lambda_{1t}
\end{align*}

The F.O.C w.r.t. $C_{2t}$ is:
\begin{align*}
\lambda_{Ct}\frac{\partial C_t}{\partial C_{2t}}=\lambda_{2t}
\end{align*}

The F.O.C w.r.t. $C_{t}$ is:
\begin{align*}
U_{Ct}=\lambda_{Ct}
\end{align*}

The F.O.C w.r.t. $K_{t+1}$ is:
\begin{align*}
\beta \lambda_{I,t+1} (1-\delta) + \beta \chi_{k,t+1} = \lambda_{It}
\end{align*}

The F.O.C w.r.t. $K_{1t}$ is:
\begin{align*}
\lambda_{1t} F_{1kt} = \chi_{kt}
\end{align*}

The F.O.C w.r.t. $K_{2t}$ is:
\begin{align*}
\lambda_{2t} F_{2kt} = \chi_{kt}
\end{align*}

The F.O.C w.r.t. $K_{Et}$ is:
\begin{align*}
\chi_{Et} F_{Ekt} = \chi_{kt}
\end{align*}

The F.O.C w.r.t. $I_{1t}$ is:
\begin{align*}
\lambda_{It} F_{I1t} = \lambda_{1t}
\end{align*}

The F.O.C w.r.t. $I_{2t}$ is:
\begin{align*}
\lambda_{It} F_{I2t} = \lambda_{2t}
\end{align*}

The F.O.C w.r.t. $L_{1t}$ is:
\begin{align*}
\lambda_{1t} F_{1lt}  = \chi_{lt}
\end{align*}

The F.O.C w.r.t. $L_{2t}$ is:
\begin{align*}
\lambda_{2t} F_{2lt} = \chi_{lt}
\end{align*}

The F.O.C w.r.t. $L_{Et}$ is:
\begin{align*}
\chi_{Et} F_{Elt} = \chi_{lt}
\end{align*}

The F.O.C w.r.t. $E_{1t}$ is:
\begin{align*}
\lambda_{1t} F_{1Et} - \lambda_{It} \mu_t \Theta^{'} - \chi_{Et}
=\sum_{j=0}^\infty \beta^j \xi_{t+j} (1-\mu_t) \frac{\partial T_{t+j}}{\partial E_t^{unc}}
\end{align*}

The F.O.C w.r.t. $E_{2t}$ is:
\begin{align*}
\lambda_{2t} F_{2Et} - \lambda_{It} \mu_t \Theta^{'} - \chi_{Et}
=\sum_{j=0}^\infty \beta^j \xi_{t+j} (1-\mu_t) \frac{\partial T_{t+j}}{\partial E_t^{unc}}
\end{align*}

The F.O.C w.r.t. $T_{t}$ is:
\begin{align*}
\lambda_{1t} \frac{\partial Y_{1t}}{\partial T_t} + \lambda_{2t} \frac{\partial Y_{2t}}{\partial T_t} + \xi_t = 0
\end{align*}

The F.O.C w.r.t. $\mu_{t}$ is:
\begin{align*}
\lambda_{It} \Theta^{'} =\sum_{j=0}^\infty \beta^j \xi_{t+j} \frac{\partial T_{t+j}}{\partial E_t^{unc}}
\end{align*}

Rearranging above equations yields:
\begin{equation}
\begin{split}
    \Theta^{'} &=F_{I1t}F_{1Et}-\frac{F_{I1t}F_{1lt}}{F_{Elt}} = F_{I2t}F_{2Et}-\frac{F_{I2t}F_{2lt}}{F_{Elt}}\\
&=(-1)\sum_{j=0}^\infty  \beta^j \bigg\{ 
\frac{\lambda_{1,t+j}}{\lambda_{It}} \frac{\partial Y_{1,t+j}}{\partial T_{t+j}} + \frac{\lambda_{2,t+j}}{\lambda_{It}} \frac{\partial Y_{2,t+j}}{\partial T_{t+j}}\bigg\}\\
&=(-1)\sum_{j=0}^\infty \beta^j \left( \underbrace {\frac{U_{C,t+j}}{\lambda_{It}} \frac{\partial C_{t+j}}{\partial C_{1,t+j}} \frac{\partial Y_{1,t+j}}{\partial T_{t+j}} \frac{\partial T_{t+j}}{\partial E_t^{unc}}}_{Impacts \quad on \quad sector \quad 1}+ \underbrace {\frac{U_{C,t+j}}{\lambda_{It}} \frac{\partial C_{t+j}}{\partial C_{2,t+j}} \frac{\partial Y_{2,t+j}}{\partial T_{t+j}}\frac{\partial T_{t+j}}{\partial E_t^{unc}}}_{Impacts \quad on\quad sector \quad 2} \right)  
\end{split}
\end{equation}
where the government’s discounting of output damages in period $t$ is governed by:
\begin{align}
\frac{\lambda_{It}}{\beta\lambda_{I,t+1}}
=F_{I1,t+1}F_{1k,t+1} +(1-\delta)=F_{I2,t+1}F_{s2,t+1} +(1-\delta)
\end{align}

\emph{Q.E.D.}

\textbf{Proof of Proposition 2}

There are two final sectors in the economy. Each sector will employ rival factors including physical capital labor, and energy for production. Production in both sectors requires nonrival factors including knowledge capital and climate capital. Thus, the production functions are given by:
\begin{equation}
\begin{split}
    & Y_{1t} = \hat{F_1}(A_{1t},T_t)K_{1t}^{\alpha}L_{1t}^{1-\alpha-\nu}E_{1t}^{\nu}\\
& Y_{2t} = \hat{F_2}(A_{2t},T_t)K_{2t}^{\alpha}L_{2t}^{1-\alpha-\nu}E_{2t}^{\nu}\    
\end{split}
\end{equation}
In addition, there is an intermediate energy production using capital and labor:
\begin{align}
E_{t} = A_{Et}K_{Et}^{\alpha_E} L_{Et}^{1-\alpha_E}
\end{align}
Note that we assume the energy sector is immune to climate change.

Then, market clearing conditions will be given by:
\begin{equation}
\begin{split}
& L_t = L_{1t}+L_{2t}+L_{et}\\
& K_t = K_{1t}+K_{2t}+K_{et}\\
& E_t = E_{1t}+E_{2t}   
\end{split}
\end{equation}

Firms in two final sectors shall decide their production plans to maximize profits:
\begin{align}
\max_{K_{it},L_{it},E_{it}} p_{it}Y_{it}-r_{t}K_{it}-w_{t}L_{it}-p_{Et}E_{it}
\end{align}

If the government implements no carbon tax, firms in energy sector will decide their production plans according to:
\begin{align}
\max_{K_{Et},L_{Et}} p_{Et}E_{t}-r_{t}K_{Et}-w_{t}L_{Et}
\end{align}

Thus, one can obtain the identity across sectors using the interest rate:
\begin{equation}
\label{eq:rental rate}
\begin{split}
r_{t} &= \alpha p_{1t}\hat{F_1}(A_{1t},T_t)\left( \frac{K_{1t}}{L_{1t}}\right)^{\alpha-1}\left( \frac{E_{1t}}{L_{1t}}\right)^{\nu} \\
&= \alpha p_{2t}\hat{F_2}(A_{2t},T_t)\left( \frac{K_{2t}}{L_{2t}}\right)^{\alpha-1}\left( \frac{E_{2t}}{L_{2t}}\right)^{\nu} \\
&= \alpha_E p_{Et} A_{Et}  \left( \frac{K_{Et}}{L_{Et}} \right)^{\alpha_E-1}   
\end{split}
\end{equation}

In a similar vein, the identity using the wage will be:
\begin{equation}
\begin{split}
w_{t} &= (1-\alpha-\nu) p_{1t}\hat{F_1}(A_{1t},T_t)\left(\frac{K_{1t}}{L_{1t}}\right)^{\alpha}\left(\frac{E_{1t}}{L_{1t}}\right)^{\nu} \\
&= (1-\alpha-\nu) p_{2t}\hat{F_2}(A_{2t},T_t)\left(\frac{K_{2t}}{L_{2t}}\right)^{\alpha}\left(\frac{E_{2t}}{L_{2t}}\right)^{\nu} \\
&= (1-\alpha_E)p_{Et} A_{Et} \left( \frac{K_{Et}}{L_{Et}} \right)^{-\alpha_E}
\end{split}
\end{equation}

Finally, the identity using the energy price is given by:
\begin{equation}
\label{eq:energy price}
\begin{split}
p_{Et} &= \nu p_{1t}\hat{F_1}(A_{1t},T_t) \left(\frac{K_{1t}}{L_{1t}}\right)^{\alpha} \left(\frac{E_{1t}}{L_{1t}}\right)^{\nu-1} \\
&= \nu p_{2t}\hat{F_2}(A_{2t},T_t) \left(\frac{K_{2t}}{L_{2t}}\right)^{\alpha} \left(\frac{E_{2t}}{L_{2t}}\right)^{\nu-1}\
\end{split}
\end{equation}

Combining the above three identities, the capital-labor ratios between three sectors should satisfy:
\begin{align}
\label{eq:capital-labor}
\frac{K_{1t}}{L_{1t}} &= \frac{K_{2t}}{L_{2t}}=\frac{K_{Et}}{L_{Et}}
\end{align}

Likewise, the energy-labor ratios between two final sectors should satisfy:
\begin{align}
\label{eq:energy-labor}
\frac{E_{1t}}{L_{1t}} &= \frac{E_{2t}}{L_{2t}}
\end{align}

Substituting Eq.(\ref{eq:capital-labor})-Eq.(\ref{eq:energy-labor}) into Eq.(\ref{eq:rental rate}) to Eq.(\ref{eq:energy price}), one can obtain the relative price of services to goods is given by:
\begin{align}
\label{eq:relative price}
\frac{p_{2t}}{p_{1t}}=\frac{\hat{F_1}(A_{1t},T_t)}{\hat{F_2}(A_{2t},T_t)}
\end{align}

Given the production productions of consumption and investment Eq.(\ref{eq:consumption}) and Eq.(\ref{eq:investment}), it is straightforward to show that the cost minimization problem for producing both yields:
\begin{align}
\frac{p_{1t}C_{1t}}{p_{2t}C_{2t}} &= \frac{\omega_c}{1-\omega_c}\left(\frac{p_{1t}}{p_{2t}}\right)^{1-\epsilon_c}\\
\frac{p_{1t}I_{1t}}{p_{2t}I_{2t}} &= \frac{\omega_I}{1-\omega_I}\left(\frac{p_{1t}}{p_{2t}}\right)^{1-\epsilon_I}
\end{align}

Substituting Eq.(\ref{eq:relative price}) into above gives:
\begin{align}
\label{eq:relative consumption}
\frac{C_{1t}}{C_{2t}} &= \frac{\omega_c}{1-\omega_c}\left(\frac{\hat{F_1}(A_{1t},T_t)}{\hat{F_2}(A_{2t},T_t)}\right)^{\epsilon_c}\\
\frac{I_{1t}}{I_{2t}} &= \frac{\omega_I}{1-\omega_I}\left(\frac{\hat{F_1}(A_{1t},T_t)}{\hat{F_2}(A_{2t},T_t)}\right)^{\epsilon_I}
\end{align}

The consumption production function Eq.(\ref{eq:consumption}) can be rewritten in two different ways:
\begin{align}
    C_t &= \left(\omega_c^{\frac{1}{\epsilon_c}} + (1-\omega_c)^{\frac{1}{\epsilon_c}}  \left(\frac{C_{2t}}{C_{1t}}\right)^{\frac{\epsilon_c-1}{\epsilon_c}} \right) ^ {\epsilon_c} C_{1t}\\
    C_t &= \left(\omega_c^{\frac{1}{\epsilon_c}} \left(\frac{C_{1t}}{C_{2t}}\right)^{\frac{\epsilon_c-1}{\epsilon_c}} + (1-\omega_c)^{\frac{1}{\epsilon_c}}  \right) ^ {\frac{\epsilon_c}{\epsilon_c-1}} C_{2t}
\end{align}

Combining Eq.(\ref{eq:relative consumption}) and above two equations, we obtain:
\begin{align}
    C_t &= \left(\omega_c^{\frac{1}{\epsilon_c}} + \frac{(1-\omega_c)}{\omega_c^{\frac{\epsilon_c-1}{\epsilon_c}}} \left(\frac{\hat{F_2}(A_{2t},T_t)}{\hat{F_1}(A_{1t},T_t)}\right)^{\frac{\epsilon_c-1}{\epsilon_c}} \right) ^ {\frac{\epsilon_c}{\epsilon_c-1}} C_{1t}\\
    C_t &= \left(\frac{\omega_c}{(1-\omega_c)^{\frac{\epsilon_c-1}{\epsilon_c}}} \left(\frac{\hat{F_1}(A_{1t},T_t)}{\hat{F_2}(A_{2t},T_t)}\right)^{\frac{\epsilon_c-1}{\epsilon_c}} + (1-\omega_c)^{\frac{1}{\epsilon_c}}  \right) ^ {\frac{\epsilon_c}{\epsilon_c-1}} C_{2t}
\end{align}

Rearranging both equations and adding together, we have:
\begin{align}
    C_t &= \left(\omega_c  \hat{F_1}(A_{1t},T_t)^{\epsilon_c-1} + (1-\omega_c) \hat{F_2}(A_{2t},T_t)^{\epsilon_c-1} \right) ^ {\frac{1}{\epsilon_c-1}} \left(\frac{C_{1t}}{\hat{F_1}(A_{1t},T_t)}+ \frac{C_{2t}}{\hat{F_2}(A_{2t},T_t)}\right)
\end{align}

In a similiar vein, for investment we have:
\begin{align}
    I_t &= A_{It} \left(\omega_I  \hat{F_1}(A_{1t},T_t)^{\epsilon_I-1} + (1-\omega_I) \hat{F_2}(A_{2t},T_t)^{\epsilon_I-1} \right) ^ {\frac{1}{\epsilon_I-1}} \left(\frac{I_{1t}}{\hat{F_1}(A_{1t},T_t)}+ \frac{I_{2t}}{\hat{F_2}(A_{2t},T_t)}\right)
\end{align}

Thus, the transformation ratio between two definitions of social cost of carbon is given by:
\begin{align}
    \frac{SCC_{CE}}{SCC_{IE}}&=\frac{\lambda_I}{\lambda_c} \notag\\
    &=\frac{F_{C1t}}{F_{I1t}}=\frac{F_{C2t}}{F_{I2t}} \notag\\
    &=\frac{\left[\omega_c  \hat{F_1}(A_{1t},T_t)^{\epsilon_c-1} + (1-\omega_c) \hat{F_2}(A_{2t},T_t)^{\epsilon_c-1} \right] ^ {\frac{1}{\epsilon_c-1}}}{A_{It} \left[\omega_I  \hat{F_1}(A_{1t},T_t)^{\epsilon_I-1} + (1-\omega_I) \hat{F_2}(A_{2t},T_t)^{\epsilon_I-1} \right] ^ {\frac{1}{\epsilon_I-1}}}
\end{align}

\emph{Q.E.D.}

\section{Calibration details}
\label{sec:appendix:two}

\begin{table}[htbp]
  \centering
  \caption{Parameters in climate module common to all models}
  \footnotesize
    \begin{tabular}{lll}
    \toprule
    Parameters    & \multicolumn{1}{l}{Value} & Sources \& notes \\
    \midrule
    \multicolumn{3}{l}{Carbon cycle equations: $\begin{pmatrix} M_t^{At}\\M_t^{Up}\\M_t^{Lo} \end{pmatrix}=
    \begin{pmatrix} \phi_{11} & \phi_{21} & 0\\\phi_{12} & \phi_{22} & \phi_{32} \\ 0 & \phi_{23} & \phi_{33} \end{pmatrix}
    \begin{pmatrix} M_{t-1}^{At}\\M_{t-1}^{Up}\\M_{t-1}^{Lo}\end{pmatrix}+
    \begin{pmatrix} E_t^{M}+E_t^{Land}\\0\\0 \end{pmatrix}$}\\
    \midrule
    $M_{2015}^{At}$ & 851   & \citet{nordhaus2017revisiting}, GtC \\
    $M_{2015}^{Up}$ & 460   & \citet{nordhaus2017revisiting}, GtC \\
    $M_{2015}^{Lo}$ & 1740  & \citet{nordhaus2017revisiting}, GtC \\
    $E_{2015}^{Land}$ & 2.6   & GtCO2/year \\
    $E_{t}^{Land}$ & $E_{2015}^{Land}(0.885)^t$ & \citet{nordhaus2017revisiting}, per five years\\
    $\phi-{11}$ & 0.88  & \citet{nordhaus2017revisiting} \\
    $\phi_{21}$ & 0.196 & \citet{nordhaus2017revisiting} \\
    $\phi_{12}$ & 0.12  & \citet{nordhaus2017revisiting} \\
    $\phi_{22}$ & 0.797 & \citet{nordhaus2017revisiting} \\
    $\phi_{23}$ & 0.007 & \citet{nordhaus2017revisiting} \\
    $\phi_{32}$ & 0.001 & \citet{nordhaus2017revisiting} \\
    $\phi_{33}$ & 0.999 & \citet{nordhaus2017revisiting} \\
    \midrule
    \multicolumn{3}{l}{Radiative forcings: $\chi_{t} = \kappa \left[ \ln \left( M_t^{At}/ M_{1750}^{At}\right) / \ln(2) \right] + \chi_{t}^{Ex}$ }\\
    \midrule
    $\kappa$ & 3.6813 &  \\
    $M_{1750}^{At}$ & 588   & \citet{nordhaus2017revisiting}, GtC \\
    $\chi_{2015}^{Ex}$ & 0.5   & \citet{nordhaus2017revisiting}, Watt/m2 \\
    $\chi_{2100}^{Ex}$ & 1     & \citet{nordhaus2017revisiting}, Watt/m2 \\
    \midrule
    \multicolumn{3}{l}{Temperature:$\begin{pmatrix} T_t\\T_t^{Lo} \end{pmatrix}= \begin{pmatrix} 1-\zeta_1\zeta_2-\zeta_1\zeta_3 & \zeta_1\zeta_3  \\ 1-\zeta_4 & \zeta_4 \end{pmatrix}
    \begin{pmatrix} T_{t-1}\\T_{t-1}^{Lo} \end{pmatrix}+
    \begin{pmatrix} \zeta_1\chi_t\\0 \end{pmatrix}$   }\\
    \midrule
    $T_0$  & 0.85  & \citet{nordhaus2017revisiting}, degree \\
    $T_0^{Lo}$ & 0.0068 & \citet{nordhaus2017revisiting}, degree \\
    $\zeta_1$ &       & \citet{nordhaus2017revisiting} \\
    $\zeta_2$ &       & \citet{nordhaus2017revisiting} \\
    $\zeta_3$ &       & \citet{nordhaus2017revisiting} \\
    $\zeta_4$ &       & \citet{nordhaus2017revisiting} \\
    \bottomrule
    \end{tabular}%
\end{table}%

\begin{table}[htbp]
  \centering
  \caption{Economic frontiers in the DICE-like model}\label{tab:frontier}
    \footnotesize
    \begin{tabular}{p{1.5cm} p{3.2cm} p{8cm}}
    \toprule
    Parameters    & \multicolumn{1}{l}{Value} & Sources \& notes \\
    \midrule
    \multicolumn{3}{l}{Preference: $U= \left[ \left(\omega_c^{\frac{1}{\epsilon_c}}  C_{1t}^{\frac{\epsilon_c-1}{\epsilon_c}} + (1-\omega_c)^{\frac{1}{\epsilon_c}}  C_{2t}^{\frac{\epsilon_c-1}{\epsilon_c}} \right) ^ {\frac{\epsilon_c}{\epsilon_c-1}} \right]^{1-\sigma} / (1-\sigma)$} \\
    \midrule
    $\sigma$ & 1.45  & \citet{nordhaus2017revisiting} \\
    $\beta$ & 0.985 & \citet{nordhaus2017revisiting}\\
    $\omega_c$ & 0.5   & Consumption expenditures are implicitly assumed to be equalized in the DICE model \\
    $\epsilon_c$ & 0.2   & Table \ref{tab:utility} \\
    \midrule
    \multicolumn{3}{l}{Final production sectors: $Y_{it} =A_{it}K_{it}^{\alpha}L_{it}^{1-\alpha-\nu}E_{it}^{\nu},\quad i\in\{1,2\}$} \\
    \midrule
    $\alpha$ & 0.3   & \citet{nordhaus2017revisiting} \\
    $\nu$ & 0.03  & \citet{golosov2014optimal} \\
    $Y_{2015}$ & 105.1774 & \citet{nordhaus2017revisiting}, trillion 2010 US\$ ;$Y_t=p_{1t}Y_{1t}+p_{2t}Y_{2t}$(note: $p_{1t}=p_{2t}=1$) \\
    $Y_{1,2015}$ & 35.3032 & \multirow{2}[0]{*}{\makecell{According to World Bank, the goods sector accounts\\ for 33.47\% in 2015, trillion 2010 US\$}} \\
    $Y_{2,2015}$ & 69.8742 &  \\
    $K_{2015}$ & 223   & \citet{nordhaus2017revisiting}, trillion 2010 US\$, \\
    $L_{2015}$ & 7403  & \citet{nordhaus2017revisiting}, million, calibrated to \citet{nordhaus2017revisiting} \\
    $E_{2015}$ & 35.85 & \citet{nordhaus2017revisiting}, GtCO2/year, carbon-based energy \\
    $K_{1,2015}$ & 70.634 & \multirow{4}[0]{*}{\makecell{Net of factors used in energy sector, factor shares\\ in two final sectors are equal to value added shares}} \\
    $K_{2,2015}$ & 139.8029 &  \\
    $L_{1,2015}$ & 2240.8038 &  \\
    $L_{2,2015}$ & 4820.9785 &  \\
    $E_{1,2015}$ & 23.8168 & \multirow{2}[0]{*}{\makecell{Energy shares in two final sectors are equal to \\value added shares}} \\
    $E_{2,2015}$ & 12.0332 &  \\
    $A_{1,2015}$ & 5.024 & $=Y_{1,2015}/(K_{1,2015}^{\alpha}L_{1,2015}^{1-\alpha-\nu}E_{1,2015}^{\nu})$ \\
    $A_{2,2015}$ & 5.024 & $=Y_{2,2015}/(K_{2,2015}^{\alpha}L_{2,2015}^{1-\alpha-\nu}E_{2,2015t}^{\nu})$ \\
    $\gamma_{1,2015}$ & 0.076  & \citet{nordhaus2017revisiting} \\
    $\gamma_{2,2015}$ & 0.076  & \citet{nordhaus2017revisiting} \\
    \midrule
    \multicolumn{3}{l}{Energy sector: $E_t=A_{Et}K_{Et}^{\alpha_E} L_{Et}^{1-\alpha_E}$} \\
    \midrule
    $K_{E,2015}$ & 12.5631 & \multirow{2}[1]{*}{\makecell{Initial shares are pinned down according to equalized\\ interest rate and wage between sectors}} \\
    $L_{E,2015}$ & 131.2177 &  \\
    $\alpha_E$ & 0.597 & \citet{barrage2020optimal}\\
    $A_{E,2015}$ & 17.9382 & $=E_{2015}/(K_{E,2015}^{\alpha_E} L_{E,2015}^{1-\alpha_E})$ \\
    $\gamma_{E,2015}$ & $(1+\gamma_{1,2015})^\frac{\alpha_E}{1-\alpha-\nu}-1$ & Assume that energy sector shares the same labor-augmenting technical change as final sector \\
    \midrule
    \multicolumn{3}{l}{Investment production: $I_t = A_{It} \left(\omega_I^{\frac{1}{\epsilon_I}}  I_{1t}^{\frac{\epsilon_I-1}{\epsilon_I}} + (1-\omega_I)^{\frac{1}{\epsilon_I}}  I_{2t}^{\frac{\epsilon_I-1}{\epsilon_I}} \right) ^ {\frac{\epsilon_I}{\epsilon_I-1}}$  } \\
    \midrule
    $\omega_c$ & 0.5   & Investment expenditures are implicitly assumed to be equalized in the DICE model \\
    $\epsilon_c$ & 0.5   & Table \ref{tab:invest}\\
    $A_{I,2015}$ & 1     & Normalized \\
    $\gamma_{I,2015}$ & 0     & Assume no investment-specific technical change \\
    \bottomrule
    \end{tabular}%
\end{table}%

\begin{table}[htbp]
  \centering
  \caption{Parameters in abatement cost function common to all models}
  \footnotesize
    \begin{tabular}{p{1.5cm} p{3.2cm} p{8cm}}
    \toprule
    Parameters    & \multicolumn{1}{l}{Value} & Sources \& notes \\
    \midrule
    \multicolumn{3}{l}{Abatement cost function: $\Theta_t(\mu_t E_t)=\frac{\overline{a}P_t^{backstop}}{1+a_t \exp(b_{0t}-b_{1t}(\mu_t E_t)^{b2})} \cdot (\mu_t E_t)$} \\
    \midrule
    $\bar{a}$ & 0.7464 & \multirow{5}[0]{*}{Minimizing the gap with DICE}\\
    $a_t$ & $0.0236+0.8881t$ & \\
    $b_{0t}$ & $7.8640-1.4858t$ & \\
    $b_{1t}$ & $1.6791-0.3157t$ & \\
    $b_{2}$ & 0.4207 & \\
    $P_{t}^{backstop}$ & $550\times (1-0.025)^{t-1} $& \citet{nordhaus2017revisiting}, per five year for a period, 2010US\$/tCO2 \\
    \bottomrule
    \end{tabular}%
  \label{tab:addlabel}%
\end{table}%

\begin{table}[htbp]
  \centering
  \caption{Homogeneous climate vulnerability in DICE-like models}\label{eq:hom par}
  \footnotesize
    \begin{tabular}{lll}
    \toprule
    Parameters    & \multicolumn{1}{l}{Value} & Sources \& notes \\
    \midrule
    \multicolumn{3}{l}{Damage function: $1-D_{i}(T_t)=1/(1+\theta_i *T_t^2),\quad i\in\{1,2\}$} \\
    \midrule
    $\theta_1$ & 0.00236 & \multirow{2}[0]{*}{\makecell{DICE implicitely assumes homogeneous climate vulnerability\\ between sectors}}\\
    $\theta_2$ & 0.00236 & \\
    \bottomrule
    \end{tabular}%
\end{table}%

\begin{table}[htbp]
  \centering
  \caption{Economic modules with heterogeneous climate vulnerability in DICE-like models}\label{tab:het par}
  \footnotesize
    \begin{tabular}{lll}
    \toprule
    Parameters    & \multicolumn{1}{l}{Value} & Sources \& notes \\
    \midrule
    \multicolumn{3}{l}{Final production sector: $Y_{it} =(1-D_{i}(T_t))A_{it}K_{it}^{\alpha}L_{it}^{1-\alpha-\nu}E_{it}^{\nu},\quad i\in\{1,2\}$} \\
    \midrule
    $Y_{1,2015}/((1-D_{1})$ & 35.3032 & Table \ref{tab:frontier} \\
    $Y_{2,2015}/((1-D_{2})$ & 69.8742 & Table \ref{tab:frontier} \\
    $K_{1,2015}$ & 70.5347 & \multirow{6}[0]{*}{In line with value added shares, adjusted by climate damages} \\
    $K_{2,2015}$ & 139.9023 &  \\
    $L_{1,2015}$ & 2237.3701 &  \\
    $L_{2,2015}$ & 4834.4122 &  \\
    $E_{1,2015}$ & 23.8337 &  \\
    $E_{2,2015}$ & 12.0163 &  \\
    $A_{1,2015}$ & 5.0311 & $=Y_{1,2015}/((1-D_{1}(T_{2015}))K_{1,2015}^{\alpha}L_{1,2015}^{1-\alpha-\nu}E_{1,2015}^{\nu})$ \\
    $A_{2,2015}$ & 5.0205 & $=Y_{2,2015}/((1-D_{1}(T_{2015}))K_{2,2015}^{\alpha}L_{2,2015}^{1-\alpha-\nu}E_{2,2015t}^{\nu})$ \\
    \midrule
    \multicolumn{3}{l}{Damage function: $1-D_{i}(T_t)=1/(1+\theta_i *T_t^2),\quad i\in\{1,2\}$} \\
    \midrule
    $\theta_1$ & 0.001414 & \multirow{2}[0]{*}{\makecell{Aggregate damage amounts to that of DICE, and the impact\\ on goods is three times that on services}}\\
    $\theta_2$ & 0.004352 & \\  
    \bottomrule
    \end{tabular}%
\end{table}%

\begin{table}[htbp]
  \centering
  \caption{Economic frontiers under structural change}\label{tab:frontier of SC}
  \footnotesize
    \begin{tabular}{p{2.3cm} p{3.2cm} p{8cm}}
    \toprule
    Parameters    & \multicolumn{1}{l}{Value} & Sources \& notes \\
    \midrule
    \multicolumn{3}{l}{Preference: $U= \left[ \left(\omega_c^{\frac{1}{\epsilon_c}}  C_{1t}^{\frac{\epsilon_c-1}{\epsilon_c}} + (1-\omega_c)^{\frac{1}{\epsilon_c}}  C_{2t}^{\frac{\epsilon_c-1}{\epsilon_c}} \right) ^ {\frac{\epsilon_c}{\epsilon_c-1}} \right]^{1-\sigma} / (1-\sigma)$} \\
    \midrule
    $\sigma$ & 1.45  & \citet{nordhaus2017revisiting} \\
    $\beta$ & 0.985 & \citet{nordhaus2017revisiting}\\
    $\omega_c$ & 0.75   & Table \ref{tab:utility} \\
    $\epsilon_c$ & 0.2   & Table\ref{tab:utility} \\
    \midrule
    \multicolumn{3}{l}{Final production sector: $Y_{it} =A_{it}K_{it}^{\alpha}L_{it}^{1-\alpha-\nu}E_{it}^{\nu},\quad i\in\{1,2\}$} \\
    \midrule
    $\alpha$ & 0.3   & \citet{nordhaus2017revisiting} \\
    $\nu$ & 0.03  & \citet{golosov2014optimal} \\
    $Y_{2015}$ & 105.1774 & \citet{nordhaus2017revisiting}, trillion 2010US\$; $Y_{2015}=p_{1,2015}Y_{1,2015}+p_{2,2015}Y_{2,2015}$, nominal production in the initial period is selected as the numéraire\\
    $p_{1,2015}Y_{1,2015}$ & 35.3032 & \multirow{2}[0]{*}{Based on value added shares in total nominal output} \\
    $p_{2,2015}Y_{2,2015}$ & 69.8742 &  \\
    $K_{2015}$ & 223   & \citet{nordhaus2017revisiting}, trillion 2010US\$ \\
    $L_{2015}$ & 7403  & \citet{nordhaus2017revisiting}, million, calibrated to \citet{nordhaus2017revisiting} \\
    $E_{2015}$ & 35.85 & \citet{nordhaus2017revisiting}, GtCO2/year, carbon-based energy \\
    $K_{1,2015}$ & 70.634 & \multirow{4}[0]{*}{\makecell{Net of factors used in energy sector, factor shares\\ in two final sectors are equal to value added shares}} \\
    $K_{2,2015}$ & 139.8029 &  \\
    $L_{1,2015}$ & 2240.8038 &  \\
    $L_{2,2015}$ & 4820.9785 &  \\
    $E_{1,2015}$ & 23.8168 & \multirow{2}[0]{*}{\makecell{Energy shares in two final sectors are equal to \\value added shares}} \\
    $E_{2,2015}$ & 12.0332 &  \\
    $p_{1,2015}A_{1,2015}$ & 5.024 & $=p_{1,2015}Y_{1,2015}/(K_{1,2015}^{\alpha}L_{1,2015}^{1-\alpha-\nu}E_{1,2015}^{\nu})$ \\
    $p_{2,2015}A_{2,2015}$ & 5.024 & $=p_{2,2015}Y_{2,2015}/(K_{2,2015}^{\alpha}L_{2,2015}^{1-\alpha-\nu}E_{2,2015t}^{\nu})$ \\
    $\gamma_{1,2015}$ & 0.1086  & \multirow{2}[0]{*}{\makecell{Matching Table \ref{tab:frontier} to obtain the same capital stock\\ level in 2100}} \\
    $\gamma_{2,2015}$ & 0.0362  & \\
    \midrule
    \multicolumn{3}{l}{Energy sector: $E_t=A_{Et}K_{Et}^{\alpha_E} L_{Et}^{1-\alpha_E}$} \\
    \midrule
    $K_{E,2015}$ & 12.5631 & \multirow{2}[1]{*}{\makecell{Initial shares are pinned down according to equalized\\ interest rate and wage between sectors}} \\
    $L_{E,2015}$ & 131.2177 &  \\
    $\alpha_E$ & 0.597 & \citet{barrage2020optimal}\\
    $A_{E,2015}$ & 17.9382 & $=E_{2015}/(K_{E,2015}^{\alpha_E} L_{E,2015}^{1-\alpha_E})$ \\
    $\gamma_{E,2015}$ & $(1+\gamma_{1,2015})^\frac{\alpha_E}{1-\alpha-\nu}-1$ & Assume that energy sector shares the same labor-augmenting technical change as final sector \\
    \midrule
    \multicolumn{3}{l}{Investment production:$I_t = A_{It} \left(\omega_I^{\frac{1}{\epsilon_I}}  I_{1t}^{\frac{\epsilon_I-1}{\epsilon_I}} + (1-\omega_I)^{\frac{1}{\epsilon_I}}  I_{2t}^{\frac{\epsilon_I-1}{\epsilon_I}} \right) ^ {\frac{\epsilon_I}{\epsilon_I-1}}$} \\
    \midrule
    $\omega_c$ & 0.43   & Table \ref{tab:invest}\\
    $\epsilon_c$ & 0.5   & Table \ref{tab:invest}\\
    $A_{I,2015}$ & 1     & Normalized \\
    $\gamma_{I,2015}$ & 0     & Assume no investment-specific technical change \\
    \bottomrule
    \end{tabular}%
\end{table}%

\begin{table}[htbp]
  \centering
  \caption{Homogeneous and heterogeneous climate vulnerability under structural change}
  \footnotesize
    \begin{tabular}{lll}
    \toprule
    Parameters  & \multicolumn{1}{l}{Value} & Sources \& notes \\
    \midrule
    \multicolumn{3}{l}{Parameters are the same as Table \ref{eq:hom par} and Table \ref{tab:het par}} \\
    \bottomrule
    \end{tabular}%
\end{table}%

\begin{table}[htbp]
  \centering
  \caption{Parameters in counterfactual scenarios under structural change}
  \footnotesize
    \begin{tabular}{lll}
    \toprule
    Parameters    & \multicolumn{1}{l}{Value} & Sources \& notes \\
    \midrule
    \multicolumn{3}{l}{Final production sector: $Y_{it} =(1-D_{i}(T_t))A_{it}K_{it}^{\alpha}L_{it}^{1-\alpha-\nu}E_{it}^{\nu},\quad i\in\{1,2\}$} \\
    \midrule
    $Y_{1,2015}/((1-D_{1})$ & 35.3032 & Table \ref{tab:frontier} \\
    $Y_{2,2015}/((1-D_{2})$ & 69.8742 & Table \ref{tab:frontier} \\
    $K_{1,2015}$ & 70.7335 & \multirow{6}[0]{*}{In line with value added shares, adjusted by climate damages} \\
    $K_{2,2015}$ & 139.7035 &  \\
    $L_{1,2015}$ & 2444.2399 &  \\
    $L_{2,2015}$ & 4827.5424 &  \\
    $E_{1,2015}$ & 23.8337 &  \\
    $E_{2,2015}$ & 12.0501 &  \\
    $p_{1,2015}A_{1,2015}$ & 5.0167 & $=p_{1,2015}Y_{1,2015}/((1-D_{1}(T_{2015}))K_{1,2015}^{\alpha}L_{1,2015}^{1-\alpha-\nu}E_{1,2015}^{\nu})$ \\
    $p_{2,2015}A_{2,2015}$ & 5.0276 & $=p_{2,2015}Y_{2,2015}/((1-D_{1}(T_{2015}))K_{2,2015}^{\alpha}L_{2,2015}^{1-\alpha-\nu}E_{2,2015t}^{\nu})$ \\
    \midrule
    \multicolumn{3}{l}{Damage function: $1-D_{i}(T_t)=1/(1+\theta_i *T_t^2),\quad i\in\{1,2\}$} \\
    \midrule
    $\theta_1$ & 0.004352 & \multirow{2}[0]{*}{\makecell{Aggregate damage amounts to that of DICE, and the impact\\ on goods is three times that on services}}\\
    $\theta_2$ & 0.001414 & \\  
    \bottomrule
    \end{tabular}%
\end{table}%

\newpage
  \bibliographystyle{elsarticle-harv} 
  \bibliography{reference}

\begin{thebibliography}{59}
\expandafter\ifx\csname natexlab\endcsname\relax\def\natexlab#1{#1}\fi
\providecommand{\url}[1]{\texttt{#1}}
\providecommand{\href}[2]{#2}
\providecommand{\path}[1]{#1}
\providecommand{\DOIprefix}{doi:}
\providecommand{\ArXivprefix}{arXiv:}
\providecommand{\URLprefix}{URL: }
\providecommand{\Pubmedprefix}{pmid:}
\providecommand{\doi}[1]{\href{http://dx.doi.org/#1}{\path{#1}}}
\providecommand{\Pubmed}[1]{\href{pmid:#1}{\path{#1}}}
\providecommand{\bibinfo}[2]{#2}
\ifx\xfnm\relax \def\xfnm[#1]{\unskip,\space#1}\fi
%Type = Article
\bibitem[{Acemoglu and Guerrieri(2008)}]{acemoglu2008capital}
\bibinfo{author}{Acemoglu, D.}, \bibinfo{author}{Guerrieri, V.}, \bibinfo{year}{2008}.
\newblock \bibinfo{title}{Capital deepening and nonbalanced economic growth}.
\newblock \bibinfo{journal}{Journal of Political Economy} \bibinfo{volume}{116}, \bibinfo{pages}{467--498}.
%Type = Article
\bibitem[{Acemoglu and Rafey(2023)}]{acemoglu2023mirage}
\bibinfo{author}{Acemoglu, D.}, \bibinfo{author}{Rafey, W.}, \bibinfo{year}{2023}.
\newblock \bibinfo{title}{Mirage on the horizon: Geoengineering and carbon taxation without commitment}.
\newblock \bibinfo{journal}{Journal of Public Economics} \bibinfo{volume}{219}, \bibinfo{pages}{104802}.
%Type = Article
\bibitem[{Alder et~al.(2022)Alder, Boppart and Muller}]{alder2022theory}
\bibinfo{author}{Alder, S.}, \bibinfo{author}{Boppart, T.}, \bibinfo{author}{Muller, A.}, \bibinfo{year}{2022}.
\newblock \bibinfo{title}{A theory of structural change that can fit the data}.
\newblock \bibinfo{journal}{American Economic Journal: Macroeconomics} \bibinfo{volume}{14}, \bibinfo{pages}{160--206}.
%Type = Article
\bibitem[{Alvarez-Cuadrado et~al.(2017)Alvarez-Cuadrado, Van~Long and Poschke}]{alvarez2017capital}
\bibinfo{author}{Alvarez-Cuadrado, F.}, \bibinfo{author}{Van~Long, N.}, \bibinfo{author}{Poschke, M.}, \bibinfo{year}{2017}.
\newblock \bibinfo{title}{Capital--labor substitution, structural change, and growth}.
\newblock \bibinfo{journal}{Theoretical Economics} \bibinfo{volume}{12}, \bibinfo{pages}{1229--1266}.
%Type = Article
\bibitem[{Arrow et~al.(2004)Arrow, Dasgupta, Goulder, Daily, Ehrlich, Heal, Levin, M{\"a}ler, Schneider, Starrett et~al.}]{arrow2004we}
\bibinfo{author}{Arrow, K.}, \bibinfo{author}{Dasgupta, P.}, \bibinfo{author}{Goulder, L.}, \bibinfo{author}{Daily, G.}, \bibinfo{author}{Ehrlich, P.}, \bibinfo{author}{Heal, G.}, \bibinfo{author}{Levin, S.}, \bibinfo{author}{M{\"a}ler, K.G.}, \bibinfo{author}{Schneider, S.}, \bibinfo{author}{Starrett, D.}, et~al., \bibinfo{year}{2004}.
\newblock \bibinfo{title}{Are we consuming too much?}
\newblock \bibinfo{journal}{Journal of Economic Perspectives} \bibinfo{volume}{18}, \bibinfo{pages}{147--172}.
%Type = Article
\bibitem[{Arrow et~al.(2012)Arrow, Dasgupta, Goulder, Mumford and Oleson}]{10.2307/26265518}
\bibinfo{author}{Arrow, K.J.}, \bibinfo{author}{Dasgupta, P.}, \bibinfo{author}{Goulder, L.H.}, \bibinfo{author}{Mumford, K.J.}, \bibinfo{author}{Oleson, K.}, \bibinfo{year}{2012}.
\newblock \bibinfo{title}{Sustainability and the measurement of wealth}.
\newblock \bibinfo{journal}{Environment and Development Economics} \bibinfo{volume}{17}, \bibinfo{pages}{317--353}.
%Type = Techreport
\bibitem[{Bakkensen and Barrage(2018)}]{NBERw24893}
\bibinfo{author}{Bakkensen, L.}, \bibinfo{author}{Barrage, L.}, \bibinfo{year}{2018}.
\newblock \bibinfo{title}{Climate Shocks, Cyclones, and Economic Growth: Bridging the Micro-Macro Gap}.
\newblock \bibinfo{type}{Working Paper} \bibinfo{number}{24893}. National Bureau of Economic Research.
\newblock \URLprefix \url{http://www.nber.org/papers/w24893}.
%Type = Article
\bibitem[{Barrage(2018)}]{barrage2018careful}
\bibinfo{author}{Barrage, L.}, \bibinfo{year}{2018}.
\newblock \bibinfo{title}{Be careful what you calibrate for: social discounting in general equilibrium}.
\newblock \bibinfo{journal}{Journal of Public Economics} \bibinfo{volume}{160}, \bibinfo{pages}{33--49}.
%Type = Article
\bibitem[{Barrage(2020)}]{barrage2020optimal}
\bibinfo{author}{Barrage, L.}, \bibinfo{year}{2020}.
\newblock \bibinfo{title}{Optimal dynamic carbon taxes in a climate--economy model with distortionary fiscal policy}.
\newblock \bibinfo{journal}{Review of Economic Studies} \bibinfo{volume}{87}, \bibinfo{pages}{1--39}.
%Type = Techreport
\bibitem[{Barrage and Nordhaus(2023)}]{dice2023}
\bibinfo{author}{Barrage, L.}, \bibinfo{author}{Nordhaus, W.D.}, \bibinfo{year}{2023}.
\newblock \bibinfo{title}{Policies, Projections, and the Social Cost of Carbon: Results from the DICE-2023 Model}.
\newblock \bibinfo{type}{Working Paper} \bibinfo{number}{31112}. National Bureau of Economic Research.
\newblock \URLprefix \url{http://www.nber.org/papers/w31112}.
%Type = Article
\bibitem[{Baumol(1967)}]{baumol1967macroeconomics}
\bibinfo{author}{Baumol, W.J.}, \bibinfo{year}{1967}.
\newblock \bibinfo{title}{Macroeconomics of unbalanced growth: The anatomy of urban crisis}.
\newblock \bibinfo{journal}{American Economic Review} \bibinfo{volume}{57}, \bibinfo{pages}{415--426}.
%Type = Article
\bibitem[{Baumol(1972)}]{baumol1972taxation}
\bibinfo{author}{Baumol, W.J.}, \bibinfo{year}{1972}.
\newblock \bibinfo{title}{On taxation and the control of externalities}.
\newblock \bibinfo{journal}{American Economic Review} \bibinfo{volume}{62}, \bibinfo{pages}{307--322}.
%Type = Article
\bibitem[{Boppart(2014)}]{boppart2014structural}
\bibinfo{author}{Boppart, T.}, \bibinfo{year}{2014}.
\newblock \bibinfo{title}{Structural change and the kaldor facts in a growth model with relative price effects and non-gorman preferences}.
\newblock \bibinfo{journal}{Econometrica} \bibinfo{volume}{82}, \bibinfo{pages}{2167--2196}.
%Type = Article
\bibitem[{Buera and Kaboski(2009)}]{buera2009can}
\bibinfo{author}{Buera, F.J.}, \bibinfo{author}{Kaboski, J.P.}, \bibinfo{year}{2009}.
\newblock \bibinfo{title}{Can traditional theories of structural change fit the data?}
\newblock \bibinfo{journal}{Journal of the European Economic Association} \bibinfo{volume}{7}, \bibinfo{pages}{469--477}.
%Type = Techreport
\bibitem[{Buera et~al.(2020)Buera, Kaboski, Mestieri and O'Connor}]{NBERw27731}
\bibinfo{author}{Buera, F.J.}, \bibinfo{author}{Kaboski, J.P.}, \bibinfo{author}{Mestieri, M.}, \bibinfo{author}{O'Connor, D.G.}, \bibinfo{year}{2020}.
\newblock \bibinfo{title}{The Stable Transformation Path}.
\newblock \bibinfo{type}{Working Paper} \bibinfo{number}{27731}. National Bureau of Economic Research.
\newblock \DOIprefix\doi{10.3386/w27731}.
%Type = Article
\bibitem[{Burke et~al.(2015)Burke, Hsiang and Miguel}]{burke2015global}
\bibinfo{author}{Burke, M.}, \bibinfo{author}{Hsiang, S.M.}, \bibinfo{author}{Miguel, E.}, \bibinfo{year}{2015}.
\newblock \bibinfo{title}{Global non-linear effect of temperature on economic production}.
\newblock \bibinfo{journal}{Nature} \bibinfo{volume}{527}, \bibinfo{pages}{235--239}.
%Type = Article
\bibitem[{Cai et~al.(2023)Cai, Brock and Xepapadeas}]{cai2023climate}
\bibinfo{author}{Cai, Y.}, \bibinfo{author}{Brock, W.}, \bibinfo{author}{Xepapadeas, A.}, \bibinfo{year}{2023}.
\newblock \bibinfo{title}{Climate change impact on economic growth: Regional climate policy under cooperation and noncooperation}.
\newblock \bibinfo{journal}{Journal of the Association of Environmental and Resource Economists} \bibinfo{volume}{10}, \bibinfo{pages}{569--605}.
%Type = Article
\bibitem[{Carleton et~al.(2022)Carleton, Jina, Delgado, Greenstone, Houser, Hsiang, Hultgren, Kopp, McCusker, Nath et~al.}]{carleton2022valuing}
\bibinfo{author}{Carleton, T.}, \bibinfo{author}{Jina, A.}, \bibinfo{author}{Delgado, M.}, \bibinfo{author}{Greenstone, M.}, \bibinfo{author}{Houser, T.}, \bibinfo{author}{Hsiang, S.}, \bibinfo{author}{Hultgren, A.}, \bibinfo{author}{Kopp, R.E.}, \bibinfo{author}{McCusker, K.E.}, \bibinfo{author}{Nath, I.}, et~al., \bibinfo{year}{2022}.
\newblock \bibinfo{title}{Valuing the global mortality consequences of climate change accounting for adaptation costs and benefits}.
\newblock \bibinfo{journal}{Quarterly Journal of Economics} \bibinfo{volume}{137}, \bibinfo{pages}{2037--2105}.
%Type = Article
\bibitem[{Casey et~al.(2021)Casey, Fried and Gibson}]{casey2021understanding}
\bibinfo{author}{Casey, G.}, \bibinfo{author}{Fried, S.}, \bibinfo{author}{Gibson, M.}, \bibinfo{year}{2021}.
\newblock \bibinfo{title}{Understanding climate damages: Consumption versus investment}.
\newblock \bibinfo{journal}{Available at SSRN 4007781} .
%Type = Article
\bibitem[{Comin et~al.(2021)Comin, Lashkari and Mestieri}]{comin2021structural}
\bibinfo{author}{Comin, D.}, \bibinfo{author}{Lashkari, D.}, \bibinfo{author}{Mestieri, M.}, \bibinfo{year}{2021}.
\newblock \bibinfo{title}{Structural change with long-run income and price effects}.
\newblock \bibinfo{journal}{Econometrica} \bibinfo{volume}{89}, \bibinfo{pages}{311--374}.
%Type = Article
\bibitem[{Cruz and Rossi-Hansberg(2023)}]{10.1093/restud/rdad042}
\bibinfo{author}{Cruz, J.L.}, \bibinfo{author}{Rossi-Hansberg, E.}, \bibinfo{year}{2023}.
\newblock \bibinfo{title}{{The economic geography of global warming}}.
\newblock \bibinfo{journal}{Review of Economic Studies} \URLprefix \url{https://doi.org/10.1093/restud/rdad042}.
%Type = Article
\bibitem[{Dell et~al.(2012)Dell, Jones and Olken}]{dell2012temperature}
\bibinfo{author}{Dell, M.}, \bibinfo{author}{Jones, B.F.}, \bibinfo{author}{Olken, B.A.}, \bibinfo{year}{2012}.
\newblock \bibinfo{title}{Temperature shocks and economic growth: Evidence from the last half century}.
\newblock \bibinfo{journal}{American Economic Journal: Macroeconomics} \bibinfo{volume}{4}, \bibinfo{pages}{66--95}.
%Type = Article
\bibitem[{Dietz and Stern(2015)}]{dietz2015endogenous}
\bibinfo{author}{Dietz, S.}, \bibinfo{author}{Stern, N.}, \bibinfo{year}{2015}.
\newblock \bibinfo{title}{Endogenous growth, convexity of damage and climate risk: How nordhaus' framework supports deep cuts in carbon emissions}.
\newblock \bibinfo{journal}{Economic Journal} \bibinfo{volume}{125}, \bibinfo{pages}{574--620}.
%Type = Article
\bibitem[{Drupp and H{\"a}nsel(2021)}]{drupp2021relative}
\bibinfo{author}{Drupp, M.A.}, \bibinfo{author}{H{\"a}nsel, M.C.}, \bibinfo{year}{2021}.
\newblock \bibinfo{title}{Relative prices and climate policy: How the scarcity of nonmarket goods drives policy evaluation}.
\newblock \bibinfo{journal}{American Economic Journal: Economic Policy} \bibinfo{volume}{13}, \bibinfo{pages}{168--201}.
%Type = Article
\bibitem[{Duarte and Restuccia(2010)}]{duarte2010role}
\bibinfo{author}{Duarte, M.}, \bibinfo{author}{Restuccia, D.}, \bibinfo{year}{2010}.
\newblock \bibinfo{title}{The role of the structural transformation in aggregate productivity}.
\newblock \bibinfo{journal}{Quarterly Journal of Economics} \bibinfo{volume}{125}, \bibinfo{pages}{129--173}.
%Type = Article
\bibitem[{Duernecker et~al.(2023)Duernecker, Herrendorf and Valentinyi}]{duernecker2017structural}
\bibinfo{author}{Duernecker, G.}, \bibinfo{author}{Herrendorf, B.}, \bibinfo{author}{Valentinyi, A.}, \bibinfo{year}{2023}.
\newblock \bibinfo{title}{{Structural Change within the Services Sector and the Future of Cost Disease}}.
\newblock \bibinfo{journal}{Journal of the European Economic Association} , \bibinfo{pages}{jvad030}.
%Type = Article
\bibitem[{Fankhauser and Tol(2005)}]{fankhauser2005climate}
\bibinfo{author}{Fankhauser, S.}, \bibinfo{author}{Tol, R.S.}, \bibinfo{year}{2005}.
\newblock \bibinfo{title}{On climate change and economic growth}.
\newblock \bibinfo{journal}{Resource and Energy Economics} \bibinfo{volume}{27}, \bibinfo{pages}{1--17}.
%Type = Article
\bibitem[{Foerster et~al.(2022)Foerster, Hornstein, Sarte and Watson}]{foerster2022aggregate}
\bibinfo{author}{Foerster, A.T.}, \bibinfo{author}{Hornstein, A.}, \bibinfo{author}{Sarte, P.D.G.}, \bibinfo{author}{Watson, M.W.}, \bibinfo{year}{2022}.
\newblock \bibinfo{title}{Aggregate implications of changing sectoral trends}.
\newblock \bibinfo{journal}{Journal of Political Economy} \bibinfo{volume}{130}, \bibinfo{pages}{3286--3333}.
%Type = Article
\bibitem[{Garc{\'\i}a-Santana et~al.(2021)Garc{\'\i}a-Santana, Pijoan-Mas and Villacorta}]{garcia2021investment}
\bibinfo{author}{Garc{\'\i}a-Santana, M.}, \bibinfo{author}{Pijoan-Mas, J.}, \bibinfo{author}{Villacorta, L.}, \bibinfo{year}{2021}.
\newblock \bibinfo{title}{Investment demand and structural change}.
\newblock \bibinfo{journal}{Econometrica} \bibinfo{volume}{89}, \bibinfo{pages}{2751--2785}.
%Type = Article
\bibitem[{Gillingham and Stock(2018)}]{gillingham2018cost}
\bibinfo{author}{Gillingham, K.}, \bibinfo{author}{Stock, J.H.}, \bibinfo{year}{2018}.
\newblock \bibinfo{title}{The cost of reducing greenhouse gas emissions}.
\newblock \bibinfo{journal}{Journal of Economic Perspectives} \bibinfo{volume}{32}, \bibinfo{pages}{53--72}.
%Type = Article
\bibitem[{Golosov et~al.(2014)Golosov, Hassler, Krusell and Tsyvinski}]{golosov2014optimal}
\bibinfo{author}{Golosov, M.}, \bibinfo{author}{Hassler, J.}, \bibinfo{author}{Krusell, P.}, \bibinfo{author}{Tsyvinski, A.}, \bibinfo{year}{2014}.
\newblock \bibinfo{title}{Optimal taxes on fossil fuel in general equilibrium}.
\newblock \bibinfo{journal}{Econometrica} \bibinfo{volume}{82}, \bibinfo{pages}{41--88}.
%Type = Article
\bibitem[{Greenwood et~al.(1997)Greenwood, Hercowitz and Krusell}]{greenwood1997long}
\bibinfo{author}{Greenwood, J.}, \bibinfo{author}{Hercowitz, Z.}, \bibinfo{author}{Krusell, P.}, \bibinfo{year}{1997}.
\newblock \bibinfo{title}{Long-run implications of investment-specific technological change}.
\newblock \bibinfo{journal}{American Economic Review} \bibinfo{volume}{87}, \bibinfo{pages}{342--362}.
\newblock \URLprefix \url{http://www.jstor.org/stable/2951349}.
%Type = Article
\bibitem[{Herrendorf et~al.(2015)Herrendorf, Herrington and Valentinyi}]{herrendorf2015sectoral}
\bibinfo{author}{Herrendorf, B.}, \bibinfo{author}{Herrington, C.}, \bibinfo{author}{Valentinyi, A.}, \bibinfo{year}{2015}.
\newblock \bibinfo{title}{Sectoral technology and structural transformation}.
\newblock \bibinfo{journal}{American Economic Journal: Macroeconomics} \bibinfo{volume}{7}, \bibinfo{pages}{104--133}.
%Type = Article
\bibitem[{Herrendorf et~al.(2013)Herrendorf, Rogerson and Valentinyi}]{herrendorf2013two}
\bibinfo{author}{Herrendorf, B.}, \bibinfo{author}{Rogerson, R.}, \bibinfo{author}{Valentinyi, A.}, \bibinfo{year}{2013}.
\newblock \bibinfo{title}{Two perspectives on preferences and structural transformation}.
\newblock \bibinfo{journal}{American Economic Review} \bibinfo{volume}{103}, \bibinfo{pages}{2752--2789}.
%Type = Incollection
\bibitem[{Herrendorf et~al.(2014)Herrendorf, Rogerson and Valentinyi}]{herrendorf2014growth}
\bibinfo{author}{Herrendorf, B.}, \bibinfo{author}{Rogerson, R.}, \bibinfo{author}{Valentinyi, A.}, \bibinfo{year}{2014}.
\newblock \bibinfo{title}{Chapter 6 - growth and structural transformation}, in: \bibinfo{editor}{Aghion, P.}, \bibinfo{editor}{Durlauf, S.N.} (Eds.), \bibinfo{booktitle}{Handbook of Economic Growth}. \bibinfo{publisher}{Elsevier}. volume~\bibinfo{volume}{2}, pp. \bibinfo{pages}{855--941}.
%Type = Article
\bibitem[{Herrendorf et~al.(2021)Herrendorf, Rogerson and Valentinyi}]{herrendorf2021structural}
\bibinfo{author}{Herrendorf, B.}, \bibinfo{author}{Rogerson, R.}, \bibinfo{author}{Valentinyi, A.}, \bibinfo{year}{2021}.
\newblock \bibinfo{title}{Structural change in investment and consumption—a unified analysis}.
\newblock \bibinfo{journal}{Review of Economic Studies} \bibinfo{volume}{88}, \bibinfo{pages}{1311--1346}.
%Type = Article
\bibitem[{Leon-Ledesma and Moro(2020)}]{leon2020rise}
\bibinfo{author}{Leon-Ledesma, M.}, \bibinfo{author}{Moro, A.}, \bibinfo{year}{2020}.
\newblock \bibinfo{title}{The rise of services and balanced growth in theory and data}.
\newblock \bibinfo{journal}{American Economic Journal: Macroeconomics} \bibinfo{volume}{12}, \bibinfo{pages}{109--146}.
%Type = Article
\bibitem[{Lucas~Jr(1988)}]{lucas1988mechanics}
\bibinfo{author}{Lucas~Jr, R.E.}, \bibinfo{year}{1988}.
\newblock \bibinfo{title}{On the mechanics of economic development}.
\newblock \bibinfo{journal}{Journal of Monetary Economics} \bibinfo{volume}{22}, \bibinfo{pages}{3--42}.
%Type = Article
\bibitem[{Moro(2015)}]{moro2015structural}
\bibinfo{author}{Moro, A.}, \bibinfo{year}{2015}.
\newblock \bibinfo{title}{Structural change, growth, and volatility}.
\newblock \bibinfo{journal}{American Economic Journal: Macroeconomics} \bibinfo{volume}{7}, \bibinfo{pages}{259--294}.
%Type = Article
\bibitem[{Newell et~al.(2021)Newell, Prest and Sexton}]{newell2021gdp}
\bibinfo{author}{Newell, R.G.}, \bibinfo{author}{Prest, B.C.}, \bibinfo{author}{Sexton, S.E.}, \bibinfo{year}{2021}.
\newblock \bibinfo{title}{The gdp-temperature relationship: Implications for climate change damages}.
\newblock \bibinfo{journal}{Journal of Environmental Economics and Management} \bibinfo{volume}{108}, \bibinfo{pages}{102445}.
%Type = Article
\bibitem[{Ngai and Pissarides(2007)}]{ngai2007structural}
\bibinfo{author}{Ngai, L.R.}, \bibinfo{author}{Pissarides, C.A.}, \bibinfo{year}{2007}.
\newblock \bibinfo{title}{Structural change in a multisector model of growth}.
\newblock \bibinfo{journal}{American Economic Review} \bibinfo{volume}{97}, \bibinfo{pages}{429--443}.
%Type = Techreport
\bibitem[{{Nobel Prize Committee}(2018)}]{Nobel2018}
\bibinfo{author}{{Nobel Prize Committee}}, \bibinfo{year}{2018}.
\newblock \bibinfo{title}{Scientific background on the Sveriges Riksbank Prize in Economic Sciences in memory of Alfred Nobel 2018: Economic growth, technological change, and climate change}.
\newblock \bibinfo{type}{Technical Report}.
%Type = Article
\bibitem[{Nordhaus(1992)}]{nordhaus1992optimal}
\bibinfo{author}{Nordhaus, W.D.}, \bibinfo{year}{1992}.
\newblock \bibinfo{title}{An optimal transition path for controlling greenhouse gases}.
\newblock \bibinfo{journal}{Science} \bibinfo{volume}{258}, \bibinfo{pages}{1315--1319}.
%Type = Article
\bibitem[{Nordhaus(2008)}]{nordhaus2008baumol}
\bibinfo{author}{Nordhaus, W.D.}, \bibinfo{year}{2008}.
\newblock \bibinfo{title}{Baumol's diseases: A macroeconomic perspective}.
\newblock \bibinfo{journal}{The B.E. Journal of Macroeconomics} \bibinfo{volume}{8}.
%Type = Article
\bibitem[{Nordhaus(2014)}]{nordhaus2014estimates}
\bibinfo{author}{Nordhaus, W.D.}, \bibinfo{year}{2014}.
\newblock \bibinfo{title}{Estimates of the social cost of carbon: Concepts and results from the dice-2013r model and alternative approaches}.
\newblock \bibinfo{journal}{Journal of the Association of Environmental and Resource Economists} \bibinfo{volume}{1}, \bibinfo{pages}{273--312}.
%Type = Article
\bibitem[{Nordhaus(2017)}]{nordhaus2017revisiting}
\bibinfo{author}{Nordhaus, W.D.}, \bibinfo{year}{2017}.
\newblock \bibinfo{title}{Revisiting the social cost of carbon}.
\newblock \bibinfo{journal}{Proceedings of the National Academy of Sciences} \bibinfo{volume}{114}, \bibinfo{pages}{1518--1523}.
%Type = Article
\bibitem[{Nordhaus and Yang(1996)}]{nordhaus1996regional}
\bibinfo{author}{Nordhaus, W.D.}, \bibinfo{author}{Yang, Z.}, \bibinfo{year}{1996}.
\newblock \bibinfo{title}{A regional dynamic general-equilibrium model of alternative climate-change strategies}.
\newblock \bibinfo{journal}{American Economic Review} , \bibinfo{pages}{741--765}.
%Type = Article
\bibitem[{Pindyck(2021)}]{pindyck2021we}
\bibinfo{author}{Pindyck, R.S.}, \bibinfo{year}{2021}.
\newblock \bibinfo{title}{What we know and don’t know about climate change, and implications for policy}.
\newblock \bibinfo{journal}{Environmental and Energy Policy and the Economy} \bibinfo{volume}{2}, \bibinfo{pages}{4--43}.
%Type = Article
\bibitem[{Romer(1986)}]{romer1986increasing}
\bibinfo{author}{Romer, P.M.}, \bibinfo{year}{1986}.
\newblock \bibinfo{title}{Increasing returns and long-run growth}.
\newblock \bibinfo{journal}{Journal of Political Economy} \bibinfo{volume}{94}, \bibinfo{pages}{1002--1037}.
%Type = Article
\bibitem[{Romer(1990)}]{romer1990endogenous}
\bibinfo{author}{Romer, P.M.}, \bibinfo{year}{1990}.
\newblock \bibinfo{title}{Endogenous technological change}.
\newblock \bibinfo{journal}{Journal of Political Economy} \bibinfo{volume}{98}, \bibinfo{pages}{S71--S102}.
%Type = Article
\bibitem[{Rudik et~al.(2021)Rudik, Lyn, Tan and Ortiz-Bobea}]{rudik_lyn_tan_ortiz-bobea_2021}
\bibinfo{author}{Rudik, I.}, \bibinfo{author}{Lyn, G.}, \bibinfo{author}{Tan, W.}, \bibinfo{author}{Ortiz-Bobea, A.}, \bibinfo{year}{2021}.
\newblock \bibinfo{title}{The economic effects of climate change in dynamic spatial equilibrium}.
\newblock \bibinfo{journal}{Working Paper} \URLprefix \url{osf.io/preprints/socarxiv/usghb}.
%Type = Article
\bibitem[{Sterner and Persson(2008)}]{doi:10.1093/reep/rem024}
\bibinfo{author}{Sterner, T.}, \bibinfo{author}{Persson, U.M.}, \bibinfo{year}{2008}.
\newblock \bibinfo{title}{An even sterner review: Introducing relative prices into the discounting debate}.
\newblock \bibinfo{journal}{Review of Environmental Economics and Policy} \bibinfo{volume}{2}, \bibinfo{pages}{61--76}.
%Type = Article
\bibitem[{Tol(1994)}]{Tol1994enpol}
\bibinfo{author}{Tol, R.S.}, \bibinfo{year}{1994}.
\newblock \bibinfo{title}{The damage costs of climate change: a note on tangibles and intangibles, applied to {DICE}}.
\newblock \bibinfo{journal}{Energy Policy} \bibinfo{volume}{22}, \bibinfo{pages}{436--438}.
%Type = Techreport
\bibitem[{Tol(2022)}]{Tol2023}
\bibinfo{author}{Tol, R.S.}, \bibinfo{year}{2022}.
\newblock \bibinfo{title}{{A meta-analysis of the total economic impact of climate change}}.
\newblock \bibinfo{type}{Working Paper Series} \bibinfo{number}{0422}. Department of Economics, University of Sussex Business School.
\newblock \URLprefix \url{https://ideas.repec.org/p/sus/susewp/0422.html}.
%Type = Article
\bibitem[{Tol(2018)}]{Tol2018}
\bibinfo{author}{Tol, R.S.J.}, \bibinfo{year}{2018}.
\newblock \bibinfo{title}{The economic impacts of climate change}.
\newblock \bibinfo{journal}{Review of Environmental Economics and Policy} \bibinfo{volume}{12}, \bibinfo{pages}{4--25}.
\newblock \URLprefix \url{http://dx.doi.org/10.1093/reep/rex027}, \DOIprefix\doi{10.1093/reep/rex027}.
%Type = Article
\bibitem[{Vogt-Schilb et~al.(2018)Vogt-Schilb, Meunier and Hallegatte}]{vogt2018starting}
\bibinfo{author}{Vogt-Schilb, A.}, \bibinfo{author}{Meunier, G.}, \bibinfo{author}{Hallegatte, S.}, \bibinfo{year}{2018}.
\newblock \bibinfo{title}{When starting with the most expensive option makes sense: Optimal timing, cost and sectoral allocation of abatement investment}.
\newblock \bibinfo{journal}{Journal of Environmental Economics and Management} \bibinfo{volume}{88}, \bibinfo{pages}{210--233}.
%Type = Article
\bibitem[{Waldinger(2022)}]{waldinger2022economic}
\bibinfo{author}{Waldinger, M.}, \bibinfo{year}{2022}.
\newblock \bibinfo{title}{The economic effects of long-term climate change: evidence from the little ice age}.
\newblock \bibinfo{journal}{Journal of Political Economy} \bibinfo{volume}{130}, \bibinfo{pages}{2275--2314}.
%Type = Article
\bibitem[{Weitzman(2010)}]{weitzman2010damages}
\bibinfo{author}{Weitzman, M.L.}, \bibinfo{year}{2010}.
\newblock \bibinfo{title}{What is the" damages function" for global warming—and what difference might it make?}
\newblock \bibinfo{journal}{Climate Change Economics} \bibinfo{volume}{1}, \bibinfo{pages}{57--69}.
%Type = Techreport
\bibitem[{Weitzman(2016)}]{NBERw22060}
\bibinfo{author}{Weitzman, M.L.}, \bibinfo{year}{2016}.
\newblock \bibinfo{title}{Some Theoretical Connections Among Wealth, Income, Sustainability, and Accounting}.
\newblock \bibinfo{type}{Working Paper} \bibinfo{number}{22060}. National Bureau of Economic Research.

\end{thebibliography}


%\begin{thebibliography}{00}



%\end{thebibliography}
\end{document}